\documentclass[aps,pra,twocolumn,groupedaddress,floatfix,longbibliography]{revtex4-2}
\usepackage{graphicx}
\usepackage{dcolumn}
\usepackage{bm}
\usepackage{amsmath}
\usepackage{xcolor}
\usepackage{braket}
\usepackage[utf8]{inputenc}
\usepackage[T1]{fontenc}
\usepackage{hyperref}
\hypersetup{
breaklinks=true,
    colorlinks=true,
    linkcolor=blue,
    citecolor=blue,
    filecolor=magenta,
    urlcolor=cyan,
}

\usepackage{amsfonts}
\usepackage{blkarray}
\usepackage{amssymb}
\usepackage{multirow}
\usepackage{mathrsfs}

\usepackage{natbib}      %%%%%%   PACKAGE NATBIB

\newcommand{\cz}{{\cal Z}}

\begin{document}

\title{
Wave-function microscopy: Derivation and anatomy of exact algebraic spinful wave functions and full
Wigner-molecular spectra of a few highly correlated rapidly rotating ultracold fermionic atoms}
\author{Constantine Yannouleas}
\email{Constantine.Yannouleas@physics.gatech.edu}
\author{Uzi Landman}
\email{Uzi.Landman@physics.gatech.edu}

\affiliation{School of Physics, Georgia Institute of Technology,
             Atlanta, Georgia 30332-0430}

%start-date: 15 January 2025}
%PRA, start-date: 11 March 2025}
%PRA-R, start-date: 04 May 2025}
\date{2 March 2025}

\begin{abstract} 
Exploring strongly correlated spinful states of few fermionic ultracold atoms in a rapidly rotating trap,
an example of which was 
recently realized for two fermionic $^6$Li atoms in an optical tweezer, we derive analytical (algebraic)
total-spin-eigenstate wavefunctions through the development and employment of a theoretical platform that
integrates exact numerical diagonalization (full configuration interaction) with symbolic language
processing. For such rapid rotations, where the atoms occupy the lowest Landau level (LLL), the obtained
algebraic expressions can address the full LLL spectrum in all its complexity, demonstrating that
their spatial, spectral, and spin characteristics manifest formation of collectively rotating and
vibrating Wigner molecules. The explicitly exhibited analytic wavefunctions (for two and three spinful
$^6$Li atoms) reproduce precisely the corresponding numerical FCI results, and they are shown
to reach beyond the limited range of applicability of previous Jastrow-type treatments. These results, and
their extension to bosonic systems, provide the impetus and analysis tools for future experimental and
theoretical simulations of larger mesoscopic systems.
\end{abstract}   

\maketitle

\section{Introduction}
\label{intr}

\subsection{Brief Historical Remarks}

{\color{black}
Following the formation in the mid 1990's of the first Bose-Einstein condensates, the surge of discovery
and advancement of methods of preparation, trapping, controlled tuning of inter-particle interactions, and
creation of optical lattices and synthetic gauge fields through atom-light interactions in different
geometries, culminated in a swell of realizations of Richard Feynman’s vision
\cite{RPF1982} for construction of
physical quantum simulators, ‘acting as nature does’ and capable of exact simulation of systems and
conditions that are otherwise, computationally or analytically excessively difficult, or plain intractable
\cite{lew22,dal11}.

Systems and phenomena that have been targeted for the application of ultra-cold-atom simulators, include
from high-T$_c$ superconductivity \cite{BD2012,GS2014}, collosal magneto-resistance \cite{Dag2007},
many-body highly-correlated phases (Luttinger-liquid and polymeric Wigner-molecule states) and transport
properties of fermions in lower-dimensional (wire-like elongated quantum dots) confinements simulating
inter-quantum-dot coupling elements \cite{AG2024}, correlated Wigner-molecule states of fermionc
atoms in quantum dots formed in lattice pockets simulating moir\'e patterns in two-dimensional van der
Waals, TMD, twisted bilayer materials \cite{wang24,yann24.2}, fractional quantum Hall effect states of
spinfull trapped fermions \cite{GS2014,joch24,yann20}, to atomic frequency resonators \cite{BOL1996},
interferometry \cite{Hol1993,Camp2003}, matter wave gyroscopes \cite{Dow1998}, and the development of
scalable quantum computers with neutral atoms \cite{And2007,Neg2011}.

Here we focus on ultra-cold atom quantum simulations of a solid-state phenomenon, i.e., the fractional
quantum Hall effect (FQHE) discovered originally experimentally \cite{FQHE82} through the use of
extremely strong magnetic fields applied to semiconductors surfaces at very low temperatures, that have
led to the discovery of novel states of electrons, often called “Laughlin quantum liquids”, which are
strongly-correlated phases described by Laughlin’s FQHE wave function \cite{laug83}. Such states of matter
occur when the lowest Landau level (LLL) -- i.e., the lowest, highly-degenerate, quantized energy level of
the electrons rotating under the influence of the Lorentz force -- is partially (fractionally) filled, namely
when the filling factor $\nu$, giving the ratio between the number of particless and the number of states in
the LLL, is expressed as $\nu = 1/m$  (integer $m$). 

In early investigations on rotated ultracold atomic (Bose) gases \cite{coop08,FET09} large filling factors
$\nu \gg 1$ were realized (slow-rotation regime) leading to the observation of quantized flux vortices
\cite{MAD2000} organized in an  Abrikosov lattice \cite{KET01,ZWIE05}, which has been shown to ‘soften’
when reaching lower filling factors, and approaching the LLL
\cite{CORN04PRL,SCHW04PRA,SCHW07PRL,POL09,BRE04}. More recently, geometric squeezing allowed observation
of a single Landau gauge wave function in the LLL \cite{ZWEI21,MUELLER22,ZWEI22}, as well as a
demonstration of  the distillation of chiral edge modes in a rapidly rotating bosonic superfluid
confined by an optical boundary \cite{ZWEI24}. The earliest experiments in the fractional quantum Hall
range ($\nu  <1$), were performed on rotating bosonic atomic clusters \cite{GEM10},
and more recently a FQHE Laughlin-type state ($\nu =1/2$) has been realized for two bosonic atoms in a
(driven) optical lattice \cite{grei23};
for completeness, we take note here that Laughlin states “made of light”
have been also observed recently in two photons experiments \cite{Clark20}. In closing this brief survey,
it is noteworthy that except from a most recent experiment where the $\nu=1/2$ Laughlin state has been
observed for a system made of two fastly rotating spinful fermionic atoms in an optical tweezer
\cite{joch24}, none of the experiments noted above explore ultracold Fermi gases at $\nu < 1$.
Such systems are the topic of this theoretical study, aiming at providing benchmark results, as well as
design and analysis tools for current
and future experiments on rapidly rotating fermionic systems of increasing size and spin complexity.
}

\subsection{Motivation of current paper}

Recent unprecedented experimental advances have enabled direct exploration of the spatial organization
of a few repelling and strongly correlated particles confined in potential traps of various symmetries in
two-dimensional architectures. In the past couple of years, pioneering publications reported such
investigations in highly dissimilar materials, as far apart as electronic charge carriers in moir\'{e}
transition metal dichalcogenide (TMD) superlattices \cite{wang24}, ultracold bosonic atoms in an
optical lattice under synthetic magnetic fields \cite{grei23}, and ultracold neutral fermionic atoms in a
single rapidly rotating harmonic trap \cite{joch24}. In the latter instance, the rotating trap served as a
quantum simulator for investigations of the properties of correlated many-body states formed when the few
trapped neutral fermions (e.g., two $^6$Li ultracold atoms \cite{joch24}),  
subject to the large rotational Coriolis forces, occupy the lowest Landau level
(LLL), thus mimicking fractional quantum Hall conditions under an applied strong magnetic field.
The counterintuitive measurements of carrier densities (for $N=2-4$ particles) within a deformable twisted
bilayer moir\'e potential pocket reported in Ref.\ \cite{wang24} were
rationalized and shown to be manifestation of the interplay between the emergent formation of a
symmetry-preserving sliding Wigner molecule (WM) and that of a symmetry-breaking pinned Wigner molecule
(PWM) \cite{yann23,yann24.1,wang24}, including the crystal-field effect from the neighboring moir\'e
pockets and the strain-induced evolution of symmetry-breaking structures \cite{yann24.2}.

For proper interpretation and understanding of the results gained via the above-noted experiments
(e.g., see Ref.\ \cite{joch24}), and for the design of future explorations, it is imperative that the
wave functions describing the strongly correlated fermions trapped in the rapidly rotating trap be
accurately determined and analyzed. This is indeed the purpose of the present paper. Here, by investigating,
in addition to the case of two fermions studied experimentally in Ref.\ \cite{joch24}, the system of
$N = 3$ ultracold fermionic atoms, we show that the highly correlated state of $N=2$ fermionic $^6$Li atoms,
experimentally prepared (and {\it in situ\/} imaged in Ref.\ \cite{joch24}) is
complimentary to that of the strongly correlated WM states in TMD moir\'e materials observed via STM
imaging in Ref.\ \cite{wang24}.

Morover, in adherence with the wave-function-dominated methodology 
pioneered by Laughlin \cite{laug83,laug90,laug99} (aptly termed as ‘Wavefunctionology’ \cite{simochap}),
we will advance the above-noted experimental efforts aimed at wave function microscopy, by deriving
lowest-Landau-level (LLL) integer-polynomial-type wave functions associated with the
energy levels (both ground and excited) of the spectra of such rapidly rotating $N=2$ and $N=3$
{\it spinful\/} ultracold neutral atoms that correspond to the manifestation of collectively rotating
\cite{yann02,yann03} and vibrating \cite{yann10} WMs. (We will use the acronyms RWM for
rotating WM and RVWM for ro-vibrating WM, respectively.) Importantly, we will validate these RVWM
integer-polynomial wave functions by demonstrating their equality (in a mathematical sense) with the
corresponding full configuration-interaction (FCI) solutions of the associated many-body Schr\"odinger
equation for a few contact-interacting fermions in the LLL.

Finally, we observe that the coincidence of the Jastrow-type Halperin/Laughlin \cite{halp83,laug90}
wave functions (used in the analysis of the experiments in Ref.\ \cite{joch24}) with those derived here
(analytically and numerically) for the RVWM, is limited merely to a
subset of zero-interaction energy (0IE) states (mainly for two atoms). Consequently, the assertion in
Ref.\ \cite{joch24} in favor of an underlying physics in line with the Halperin/Laughlin approach is
deemed incomplete. We foresee that future experimental investigations on rapidly-rotating three-fermion
systems, in addition to the case of two fermions studied experimentally in Ref.\ \cite{joch24},
will compellingly support the underlying WM physics presented here.

\subsection{Plan of the paper}

\begin{figure}[t]
\centering\includegraphics[width=7.8cm]{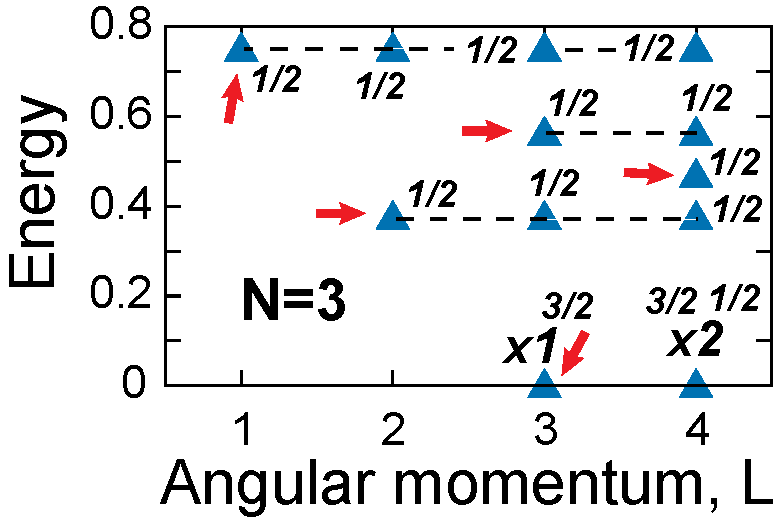}
\caption{%%%%
FCI spectrum for $N=3$ spinful contact-interacting LLL fermions with total-spin projection $S_z=1/2$.
The total spin $S=1/2$ or $S=3/2$ for each state is denoted next
to the corresponding triangular symbol. The symbol $xn$ denotes an $n$-degenerate zero-interaction
energy state. States that relate through a center-of-mass translation lie on a horizontal dashed line.
States marked by a red arrow are translationally invariant.
Energies are in units of $g/(\pi \Lambda^2)$.
}
\label{spec33}
\end{figure}

The outline of the paper is detailed in the following. In the next section (Sec.\ \ref{mbh}), we state
briefly the Hamiltonian of the contact-interacting ultracold atoms in the LLL
(for details, see Appendix \ref{appd}), and show in Figs.\
\ref{spec32} and \ref{spec33} the associated spectra in the LLL (for angular momenta $L \leq 4$),
obtained through FCI calculations for two and three ultracold fermionic atoms, respectively. The
latter case ($N=3$) manifests a level of complexity and intricacy calling for future experimental
investigation and analysis using the wave functions developed in this paper. Derivation of the analytic
wave function of the purely rotating Wigner molecule for $N=2$ spinfull fermions (or two scalar bosons)
is given in Sec.\ \ref{anan2}, along with the two-body correlation distribution corresponding to the
singlet zero-interaction-energy state at $L=2$, illustrating the spatial antipodal configuration of the
RWM. In Sec.\ \ref{anan3}, we describe the purely RWM analytic wave function for $N=3$ spinfull fermions,
along with Fig.\ \ref{3bodycsm} that exhibits the three-body correlation distribution for the ground
state (at $L=2$), revealing the equilateral triangular shape of the three fermionic RWM system.
The procedure followed for deriving the analytic wave functions, through the processing of the FCI results
with the assistance of symbolic language, is described in Sec.\ \ref{fcian}, which is followed by
Sec.\ \ref{a2} where we expose the algebraic wave functions obtained via FCI calculations for $N=3$
fermions at $L=3$. In Sec.\ \ref{halpp}, we contrast the derived (FCI-assisted), zero-interaction-energy,
analytic wave functions with the trial Halperin \cite{halp83} wave functions for $N=3$ (at $L=4$)
fermions in the LLL. This includes both comparison of the differing algebraically forms of the wave
functions considered here, as well as through analysis (shown in Fig.\ \ref{zeroes}) of the patterns of
zeroes associated with the corresponding wave functions. We conclude in Sec.\ \ref{concf}. Additional
information about the derived algebraic (analytic) wave functions for the full spectrum of the $N = 2$
fermion case and for the $N=3$ at $L=4$ case are given in Appendices \ref{appa} and \ref{appb},
respectively, with a brief note about the FCI method included in Appendix \ref{appc}.

%*********************** begin figure 1 *****************
\begin{figure}[t]
\centering\includegraphics[width=7.8cm]{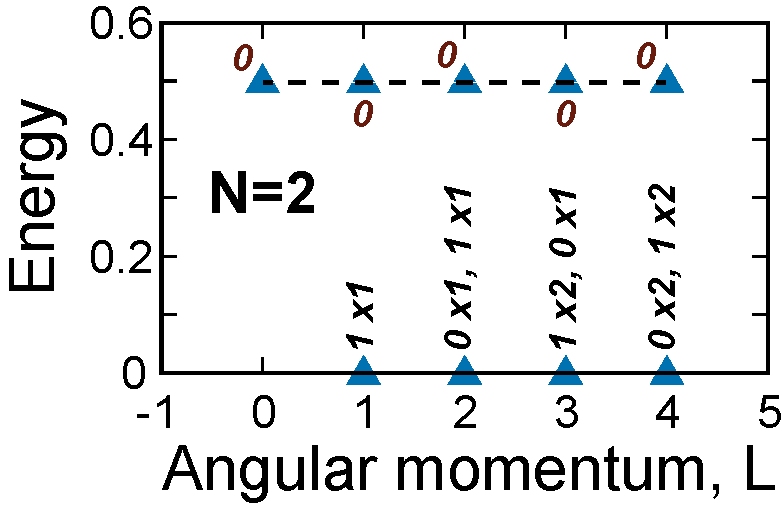}
\caption{%%%%
FCI spectrum for $N=2$ spinful contact-interacting LLL fermions with total-spin projection $S_z=0$.
The total spin $S=0$ (singlet) or $S=1$ (triplet) for each state is denoted next
to the corresponding triangular symbol. The symbol $xn$ denotes an $n$-degenerate zero-interaction
energy state. States that relate through a center-of-mass translation lie on a horizontal dashed line.
Energies are in units of $g/(\pi \Lambda^2)$.
}
\label{spec32}
\end{figure}
%*********************** end figure 1 *****************

%*********************** begin figure 2 *****************
%*********************** end figure 2 *****************

\section{The many-body Hamiltonian in the LLL and the associated spectra for $N=2$ and $N=3$}
\label{mbh}

{\color{black}
Requiring a very strong confinement of the harmonic trap
along the axis of rotation ($\hbar \omega_{z} >> \hbar \omega_{\bot}$),
freezes out the many body dynamics in the $z$-dimension, and the wavefunction
along this direction can be assumed to be permanently in the corresponding
oscillator ground state.
We are thus left with an effectively 2D system. For such a setup,
the Hamiltonian for $N$ atoms of mass $M$ in a harmonic trap $(\omega_\bot)$
rotating at angular frequency $\Omega \hat{\bf z}$ is given by:
\begin{equation}
H=\sum_{i=1}^N
\left( \frac{{\bf p}_{i}^{2}}{2M}+\frac{1}{2}
M \omega_{\bot}^{2} {\bf r}_{i}^{2} \right) -\Omega {\cal L}
+\sum_{i < j}^{N} v({\bf r}_{i}-{\bf r}_{j}).
\label{3d_hamiltonian_11}
\end{equation}
Here ${\cal L}=-\hbar L=\sum_{i=1}^N \hat{{\bf z}}\!\cdot{\bf r}_i
\times {\bf p}_{i}$ is the total angular momentum perpendicular to the $x-y$ plane;
$ {\bf r } \; =\; (x,y)$ and $ {\bf p } \; =( p_x,p_y )$ represent the single-particle position
and linear momentum in the $x-y$ plane, and $\omega_{\bot}$ is the frequency of the 2D harmonic trap.

The above  Hamiltonian can be shown for judiciously chosen conditions (applicable to the large majority of
experiments in this field) to take the form
\begin{equation}
H^\prime_{\text{LLL}}=N \hbar \omega_\bot + \hbar(\omega_{\bot}-\Omega) L
+\sum_{i < j}^{N} v({\bf r}_{i}-{\bf r}_{j}).
\label{hlll22}
\end{equation}

In the LLL, $\omega_{\bot}=\Omega$, and the kinetic energy is suppressed and thus
the many-body Hamiltonian describing ultracold neutral
atoms can be approximated with only the term containing the two-body contact interaction
\cite{popp04,yann06,coop08,hazz08,palm20,yann20,yann21,%%%%%%
[{Based on FCI calculations, RWMs for scalar bosons in the LLL were predicted in~}][{}]yann07.2},
i.e.,
\begin{equation}
H_{\rm LLL} = (g/\Lambda^2) \sum_{i < j}^{N} \delta^2(z_i-z_j),
\label{hlll}
\end{equation}
where $\Lambda=\sqrt{\hbar/(M\omega_\bot)}$, with $M$ being the mass of the atom and $\omega_\bot$ the
frequency of the rotating trap; in Eq.\ \ref{hlll}, we reverted to the common use of particle coordinates
in the 2D complex plane where the location of particle $j$ is given as $z_j = x_j + i y_j$, and the
square of the particle's distance from the origin is $z_j z^*_j$. This notation will be used throughout
the rest of the paper. $g$ is the strength of the repulsive contact interaction. For a detailed
explanation why the interaction term dominates over the kinetic energy contributions in the LLL,
see Appendix \ref{appd}.
}

Our methodoly, which integrates both numerical (e.g., fortran) and symbolic (see, e.g.,
Ref.\ \cite{math24}) languages, consists of three steps: (1) FCI numerical diagonalization of the 
Hamiltonian matrix problem associated with $H_{\rm LLL}$. (2) Analysis of the exact numerical FCI wave
functions $\Phi^{\rm CI}$ using symbolic scripts leading to determination of the
corresponding exact FCI analytical wave functions, culminating with step (3) where the latter are
connected (and vice versa) to the independently derived analytic wave functions $\Phi^{\rm RWM}$ from the
theory of rotating Wigner molecules.

The FCI spectra of the $H_{\rm LLL}$ Hamiltonian for $N=2$ and $N=3$ spinful fermions as a function of the
total angular momentum $L$ are presented in Figs.\ \ref{spec32} and \ref{spec33}, respectively.
An inspection of these figures provides testimony
to the high degree of complexity involved. Our methodology is able to provide algebraic expressions for the
full variety of all the states in such LLL spectra. Given the focus of the current experimental efforts on
assemblies of a few fermions \cite{joch24}, in this paper, we will present as an example the algebraic
expressions for $N=2$ and $N=3$ with angular momenta $L \leq 4$; the complexity uncovered here for the $N=3$
$^6$Li atoms, compared to the experimentally investigated $N=2$ case \cite{joch24}, positions the former
($N=3$) as a prime future experimental challenge.

Before proceeding with discussing the FCI wave functions and their algebraic counterparts, we
first derive in the next two sections the purely RWM analytic wave functions for two and three particles.
The purely RWM wave functions will be augmented in the following [see, e.g., Eq.\ (\ref{rwm3l2n2}) below,
Sec.\ \ref{a2}, and Appendix \ref{appb}] using  homogeneous polynomials describing vibrational
excitation modes.

\section{Derivation of the purely RWM analytic wave functions for two spinfull fermions or two
scalar bosons}
\label{anan2}

We generalize earlier analogous derivations for the case of any-$N$ fully spin
polarized electrons \cite{yann02} or scalar bosons \cite{yann10}. These derivations start naturally
from a straightforward wave function that describes through displaced Gaussians a {\it rotationally
pinned\/} WM and proceed through restoration of the broken circular symmetry via projection
techniques \cite{yann07}. Namely,

(I) At a {\it first step\/}, the broken-symmetry wave function of a fermionic $N=2$ {\it rotationally
pinned\/} WM (referred also as static WM) can be written as
\begin{align}
\Psi^{\rm PWM}_{\pm}(z_1,z_2) (\alpha(1)\beta(2) \mp \alpha(2)\beta(1))
\label{pinnwm}
\end{align}
where $\alpha$, $\beta$ denote a spin up and a spin down, respectively, and the space part is given by
symmetric and antisymmetric combinations
\begin{align}
\Psi^{\rm PWM}_{\pm}(z_1,z_2)=u(z_1,Z_1)u(z_2,Z_2) \pm u(z_2,Z_1)u(z_1,Z_2), 
\label{psi}
\end{align}
with
\begin{align}
u(z,Z_j) = \frac{1}{\sqrt{\pi}}
 \exp[-|z-Z_j|^2/2] \exp[-i (xY_j-yX_j)],
\label{gaus}
\end{align}
being a displaced Gaussian function in the LLL localized at the position $Z_j$; $z_j=x_j+iy_j=$ and
$Z_j=X_j+iY_j=R e^{i\phi_j}$. We impose $\phi_1=0$ and $\phi_2=\pi$,
so that the two {\it pinned\/} particles are in an antipodal configuration.
Lengths are in units of $\Lambda$. The phase factor is due to the gauge invariance associated with
the rotation of the trap. For two fermions, $\Psi_{+}$ is associated with a total-spin singlet state and
$\Psi_{-}$ with a triplet state. For two scalar bosons, only the symmetric $\Psi_{+}$ needs to be
considered. 

The localized orbital $u(z,Z)$ can be expanded in a series over the complete set of
zero-node Fock-Darwin \cite{yann07,fock,darw} single-particle wave functions
\begin{equation}
\psi_{l_i}(z) = \frac{ z^{l_i} } { \sqrt{ \pi {l_i}!} } \exp(-zz^*/2),
\label{psilll}
\end{equation}
with $l_i \geq 0$. One gets [see Appendix A in Ref.\ \cite{yann06.3}]
\begin{equation}
u(z,Z)=\sum_{l=0}^{\infty} C_l(Z) \psi_l(z),
\label{uexp}
\end{equation}
with 
\begin{equation}
C_l(Z)=(Z^*)^l \exp(-ZZ^*/2)/\sqrt{l!}
\label{clz}
\end{equation} 
for $Z \neq 0$. Naturally, $C_0(0)=1$ and $C_{l>0}(0)=0$.
Then the following expansion (within a proportionality constant) is obtained
\begin{align}
\Psi&^{\rm PWM}_{\pm}(z_1,z_2) =  \nonumber \\
& e^{-R^2} \sum_{l_1=0,l_2=0}^{\infty}
\frac{ (-)^{l_2}R^{l_1+l_2} } {l_1! l_2!}
 (z_1^{l_1} z_2^{l_2} \pm  z_2^{l_1} z_1^{l_2}) |0\rangle.
\label{exppsi}
\end{align}

In Eq.\ (\ref{exppsi}), the common factor $|0\rangle$ represents the product $\psi_0(z_1)\psi_0(z_2)$ of
Gaussians defined in Eq.\ (\ref{psilll}). To simplify the notation, this trivial factor will be omitted
below.

(II) {\it Second step:\/}
The pinned wave functions $\Psi^{\rm PWM}_\pm (z_1,z_2)$ break the rotational symmetry and thus
they are not eigenstates of the total angular momentum $\hbar \hat{L}=\hbar \sum_{j=1}^2 \hat{l}_j$.
However, one can restore \cite{yann02,yann04,yann07,shei21} the
rotational symmetry by applying onto $\Psi^{\rm PWM}_\pm (z_1,z_2)$ the projection operator
\begin{equation}
{\cal P}_L \equiv \frac{1}{2\pi} \int_0^{2\pi} d\gamma
e^{i\gamma(\hat{L} - L)},
\label{projl}
\end{equation}
where $\hbar L$ are the eigenvalues of the total angular momentum.

When applied onto $\Psi^{\rm PWM}_\pm (z_1,z_2)$, the projection operator ${\cal P}_L$ acts as
a Kronecker delta: from the unrestricted sum in Eq.\ (\ref{exppsi}) it picks
up only those terms having a given total angular momentum $L$ (we
drop the constant prefactor $\hbar$ when referring to angular momenta).
The spatial component of the RWM wave function, $\Psi^{\rm RWM}_{N=2}(L) ={\cal P}_L \Psi^{\rm PWM}_\pm$
(identified with the outcome of the projection \cite{yann02,yann04,yann07}), is found to be (within a
proportionality constant)
\begin{align}
\Psi^{\rm RWM}_{N=2}(L) \propto (z_1-z_2)^L,
\label{rwm2}  
\end{align}
where $L=even$ correlates with a fermionic spin singlet or a bosonic scalar state, while
$L=odd$ correlates with a fermionic spin triplet state.

%*********************** begin figure 3 *****************
\begin{figure}[t]
\centering\includegraphics[width=8.0cm]{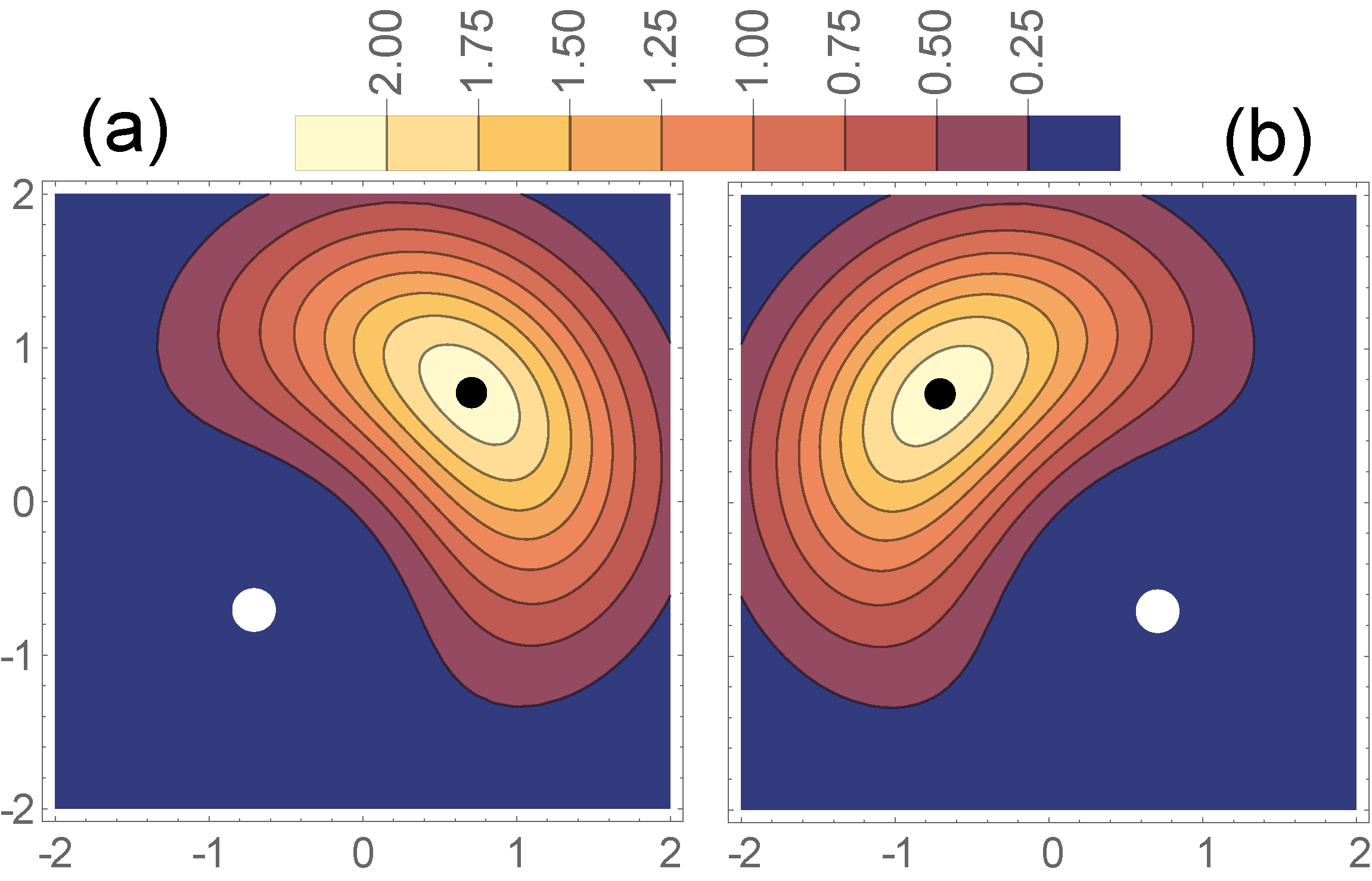}
\caption{%%%%
The two-body correlation distribution associated with the singlet 0IE state of $N=2$ LLL fermions at
$L=2$; see Eq.\ (\ref{p2l2n1}). The fixed fermion is at positions (a) $z0_1=R e^{ -3\pi i/4}$ and
(b) $z0_1=R e^{-\pi i/4}$, highlighted by white solid dots. The maximum of distribution is marked by a
black solid dot. The two dots in each frame are antipodal when $R=1 \Lambda$. Lengths in units
of $\Lambda$.
}
\label{2bodycsm}
\end{figure}
%*********************** end figure 3 *****************

The states (\ref{rwm2}) are apparently translationally invariant (TI). They are also zero-interaction
energy (0IE) states \cite{yann21} and they lie along the $x$-axis in the FCI spectra displayed in Fig.\
\ref{spec32}. The states (\ref{rwm2}) account only for a small part of the spectrum. Indeed, there are
additional 0IE states (see the degeneracies specified in  Fig.\ \ref{spec32}), as well as excitations
with non-zero energy. As an example, from the FCI analysis (see below),
we find that the second 0IE state at $L=2$ (a triplet with $S_z=0$) is $\propto (z_1^2-z_2^2)$.
(For the rest of the wave functions of the $N=2$ spectrum, see Appendix \ref{appa}.

Although the single-particle densities (first-order correlations) of the RWM wave functions are
circularly symmetric, one can visualize the intrinsic molecular configuration of a RWM by plotting
higher-order correlations. As a simple example for $N=2$, we investigate here the spin-unresolved
two-body correlations of the state at $L=2$ defined by Eq.\ (\ref{rwm2}), namely
\begin{align}
  {\cal P} (z0_1; x,y) =  (z_1-z_2)^2 (z^*_1-z^*_2)^2 e^{-\sum_{i=1}^2 z_i z_i^*},
\label{p2l2n1}
\end{align}
where one fixes one particle at point $z0_1$ and inquires about the position ($z_2=x+i y$)
of the second fermion; the star denotes complex conjugation.

In Fig.\ \ref{2bodycsm} we display the ${\cal P} (z0_1; x,y)$ defined above for two fixed points,
designated by white dots, at (a) $z0_1=R e^{ -3\pi i/4}$ and (b) $z0_1=R e^{-\pi i/4}$. 
One sees that the positions (black solid dots) associated with the maximum probability for finding the
second fermion are antipodal to the white dots when $R=1.0 \Lambda$, 
substantiating thus the physical picture of a rotating Wigner molecule.

We further note that the states (\ref{rwm2}) coincide with the Jastrow-type ones, referred to as the
``special quantum liquid'' topological Halperin/Laughlin states \cite{halp83,laug90}. However, most recent
experimental work \cite{joch24} has explicitly measured antipodal two-body quantum correlations in
agreement with the distributions in Fig.\ \ref{2bodycsm}, confirming thus the RWM underlying physics.

%*********************** begin figure 4 *****************
\begin{figure}[t]
\centering\includegraphics[width=8.0cm]{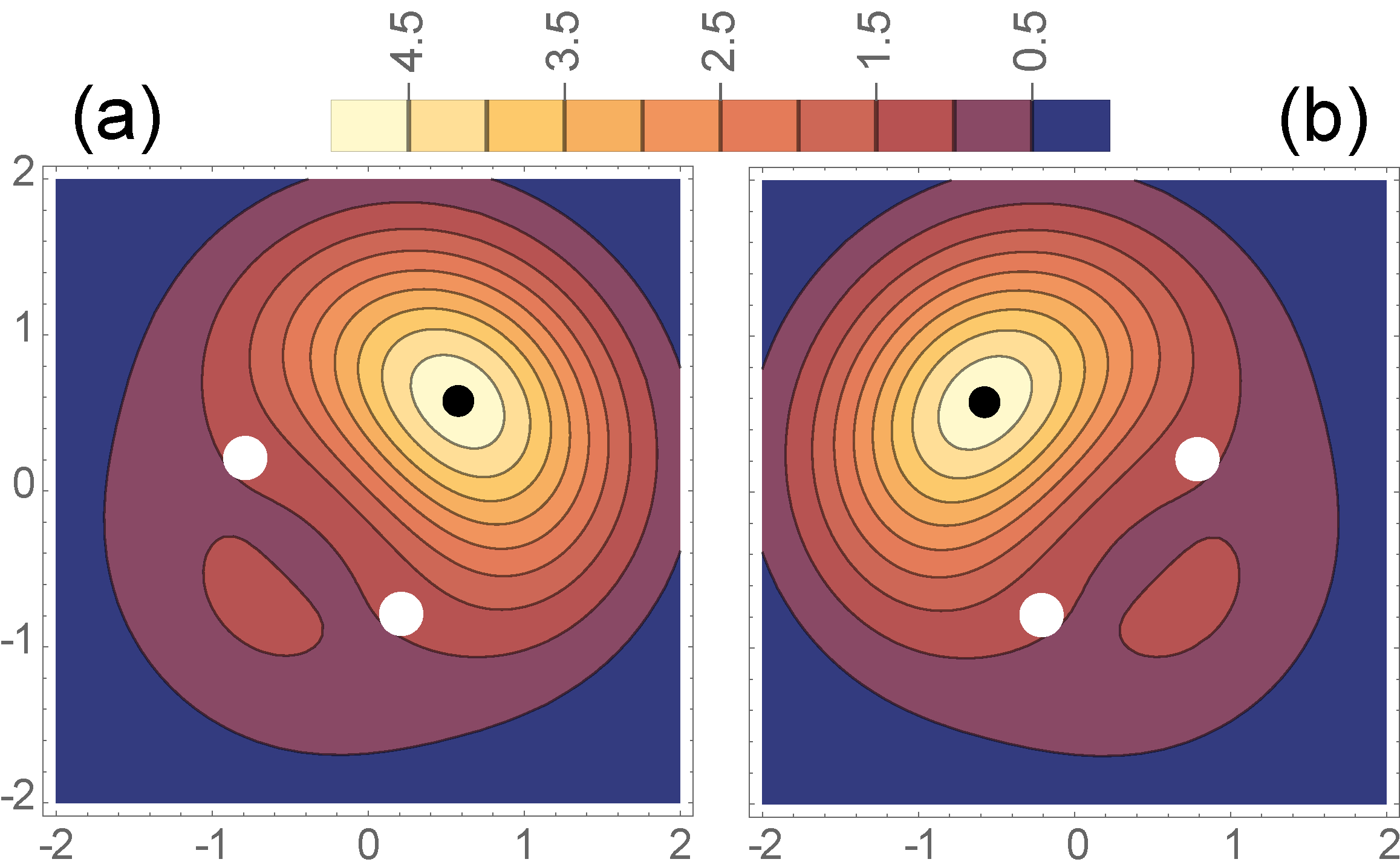}
\caption{%%%%
The three-body correlation distribution associated with the TI ground state of $N=3$ LLL fermions at
$L=2$ with spin ($S=1/2$, $S_z=1/2$); see Eq.\ (\ref{p3l2n1}). The two fixed fermions are at
positions (a) ($z0_1=R e^{ 11\pi i/12}$, $z0_2=R e^{- 5\pi i/12}$) and (b) ($z0_1=R e^{17\pi i/12}$,
$z0_2=R e^{\pi i/12}$), highlighted by white dots. The maximum of distribution is marked by a black solid
dot. The three dots in each frame are at the apices of an equilateral triangle when $R=0.816 \Lambda$.
Lengths in units of $\Lambda$.
}
\label{3bodycsm}
\end{figure}
%*********************** end figure 4 *****************

\section{Purely RWM analytic wave functions for three spinfull fermions}
\label{anan3}

The three spin eigenfunctions $\chi(S,S_z=1/2)$ associated with three fermions in
a planar equilateral trigonal configuration are given by \cite{yann03.2}
\begin{align}
& \chi(1/2,1/2;1)=(\cz_{123} + e^{2\pi i/3}\cz_{231} + e^{-2\pi i/3}\cz_{312})/\sqrt{3} \nonumber \\
& \chi(1/2,1/2;2)=(\cz_{123} + e^{-2\pi i/3}\cz_{231} + e^{2\pi i/3}\cz_{312})/\sqrt{3} \nonumber \\
& \chi(3/2,1/2)=(\cz_{123} + \cz_{231} + \cz_{312})/\sqrt{3},
\label{chi3}
\end{align}
where the spin primitives are
\begin{align}
\cz_{ijk}=\alpha(i)\alpha(j)\beta(k),
\end{align}
where $\alpha$ and $\beta$ denote up and down spins,

Three straightforward {\it pinned\/} WM wave functions, $\Phi^{\rm PWM}_{N=3}$, are constructed by
assuming displaced Gaussians $u(z,Z_j)$ [see Eq.\ (\ref{gaus})] at the apices
$Z_j=R e^{2 j \pi i/3}$, $j=0,1,2$ of an equilateral triangle, and by replacing the spin primitives
$\cz_{ijk}$ in the three $\chi$'s in Eq.\ (\ref{chi3}) by the associated Slater determinants
\begin{align}
  & {\cal D}_{ijk} = \nonumber \\
  & {\rm Det}[ u(z_i,Z_1)\alpha(i), u(z_j,Z_2)\alpha(j), u(z_k,Z_3)\beta(k) ]/ \sqrt{6},
\label{dijk}
\end{align}
where the determinants are denoted through a listing of the diagonal elements.

To obtain the {\it rotating\/} WM wave functions, $\Phi^{\rm RWM}_{ijk}(L)$, one subsequently expands
the Slater determinants, as well as the displaced Gaussians, according to Eqs.\ (\ref{uexp}) and
(\ref{clz}), and then carries out the angular momentum projection
\cite{yann02,yann07,yann06.3,shei21} by extracting the
coefficients of the powers $R^L$ ($L=l_1+l_2+l_3$).
Using symbolic scripts to sum the terms in these coefficients, one finally
obtains the compact expression in Eqs.\ (\ref{rwm3}) and (\ref{psi3}).

We note that, in accordance with well known results for the coupling between total spin and angular
momenta, each spin eigenfunction $\chi$ in Eq.\ (\ref{chi3}) is associated with different $L$'s ($L>0$)
in the compact expression (\ref{psi3}). Namely, the top $\chi(1/2,1/2;1)$ is associated with
$L=3n+1$, the middle $\chi(1/2,1/2;2)$ with $L=3n+2$, and the bottom $\chi(3/2,1/2)$ with $L=3n$,
$n$ being a nonnegative integer.

For $N=3$, the RWM wave functions with $S_z=1/2$ are given by the expression (for an outline of the
derivation, see above)
\begin{align}
\Phi^{\rm RWM}_{N=3}(L)=\sum_{ijk} \Psi^{\rm RWM}_{ijk}(L) \cz_{ijk}.
\label{rwm3}  
\end{align}
The symbolic sum variable $ijk$ in Eq.\ (\ref{rwm3}) runs over the three cyclic permutations of \{1,2,3\},
and the space parts are given by
\begin{align}
\Psi^{\rm RWM}_{ijk}(L) \propto i[(z_{i+j-k} - i z_{i-j})^L - (z_{i+j-k} + i z_{i-j})^L],
\label{psi3}
\end{align}
with $z_{i+j-k}=\sqrt{2/3}((z_i+z_j)/2-z_k)$ and $z_{i-j}=(z_i-z_j)/\sqrt{2}$ being three-particle
Jacobi coordinates, and $L$ the angular momentum.

%------------------------------------------ begin table I -------
\begin{table}[t]
\caption{\label{tn2l5n1}
The numerical FCI coefficients, $c_{\rm CI}(I)$, in the CI expansion of the 0IE LLL
state, and the corresponding extracted algebraic ones, $c_{\rm alg}(I)$,
for $N=2$ fermions with total angular momentum $L=5$ and ($S=1$, $S_z=0$).
The Slater determinants ${\cal D}_I$ are specified through the set of single-particle angular momenta
$(l_1,l_2)$ associated with up ($l_1$) and down ($l_2$) spins.
}
\begin{ruledtabular}
\begin{tabular}{rcrc}
$I$ & $c_{\rm CI} (I)$ & $c_{\rm alg} (I)$ &
$(l_1\uparrow,l_2\downarrow)$  \\ \hline
1 & 0.176777   & $ \sqrt{1/32}   $ & (0,5) \\
2 & -0.395285  & $ -\sqrt{5/32}  $ & (1,4) \\
3 & 0.559017   & $ \sqrt{10/32}  $ & (2,3) \\
4 & -0.559017  & $ -\sqrt{10/32} $ & (3,2) \\
5 & 0.395285   & $  \sqrt{5/32}  $ & (4,1) \\
6 & -0.176777  & $ -\sqrt{1/32}  $ & (5,0) \\
\end{tabular}
\end{ruledtabular}
\end{table}
%------------------------------------------ end table I --------------------

As simple examples, we analyze the single state for $L=1$ and the two states for $L=2$ whose FCI
energies are depicted in Fig.\ \ref{spec33}.

From the FCI analysis (see below), we find that the single state with $S=1/2$ at $L=1$ equals 
\begin{align}
\Phi^{\rm RWM}_{N=3}(L=1) \propto \sum_{ijk} (z_i-z_j) \cz_{ijk}.
\label{rwm3l1}
\end{align}

In addition, from the FCI analysis (see below), we find that the ground state (with $S=S_z=1/2$) at
$L=2$ is equal to
\begin{align}
\Phi^{\rm RWM}_{N=3} (L=2) \propto 
  \sum_{ijk} \Psi^{\rm RWM}_{ijk} (L=2) \cz_{ijk},
\label{rwm3l2n1}
\end{align}
where the space part is given by
\begin{align}
\Psi^{\rm RWM}_{ijk} (L=2)=(z_i-z_j) (z_i+z_j-2 z_k).
\label{psi3l2n1}
\end{align}

We investigate here the spin-unresolved three-body correlations of $\Phi^{\rm RWM}_{N=3}(L=2)$
given by
\begin{align}
  & {\cal P} (z0_1,z0_2; x,y) =  \nonumber \\
  & \sum_{ijk} \Psi^{\rm RWM}_{ijk}(L=2) \Psi^{\rm RWM*}_{ijk}(L=2)
  e^{-\sum_{i=1}^3 z_i z_i^*},
\label{p3l2n1}  
\end{align}
where one fixes two particles at points $z0_1$ and $z0_2$ and inquires about the position ($z_3=x+i y$)
of the third fermion.

In Fig.\ \ref{3bodycsm}, we display the ${\cal P} (z0_1,z0_2; x,y)$ defined above for
two pairs of fixed fermions, highlighted by white dots, at positions (a)
($z0_1=R e^{ 11\pi i/12}$, $z0_2=R e^{- 5\pi i/12}$) and (b) ($z0_1=R e^{17\pi i/12}$,
$z0_2=R e^{\pi i/12}$). One sees that the positions (black solid dots) associated with the maximum
probability for finding the third fermion form with the white dots equilateral triangles when
$R=0.816 \Lambda$, thus substantiating the physical picture of a rotating Wigner molecule.

For the excited state for $N=3$ (with $S=S_z=1/2$) at $L=2$, the FCI analysis (see below) shows that
it equals
\begin{align}
& \sum_{ijk} (z_i-z_j)(z_1+z_2+z_3) \cz_{ijk} \propto \nonumber \\
& \Phi^{\rm RWM}_{N=3} (L=1) z_{\rm c.o.m.}^{N=3},
\label{rwm3l2n2}
\end{align}
where $z_{\rm c.o.m.}=\sum_{i=1}^N z_i/N$ is the center of mass.

We note that the factor $z_{\rm c.o.m.}$,
{\color{black}
which emerges naturally (with no prior attempt at separation
of the particles' center-of-mass and relative motion degrees of freedom) in the course of the exact
diagonalization (i.e., FCI) of the microscopic Hamiltoninan, and in the subsequent derivation of the
corresponding analytical wave function,}
is a structural earmark of all states on a horizontal dashed line in the LLL spectrum. Indeed, as was
ascertained through our detailed inspections, each integer polynomial describing a state on a
horizontal dashed line at a given $L$ (see Fig.\ \ref{spec33}, and also Fig.\ \ref{spec32}) equals
that of the previous state at $L-1$ times the factor $z_{\rm c.o.m.}$, apart from the original states
that are translationally invariant marked by a red arrow (in Fig.\ \ref{spec33}).
This factor $z_{\rm c.o.m.}$ represents center-of-mass vibrations which in the LLL case degenerate into
simple translations. 

In Sec.\ \ref{halpp},
we discuss in detail that the rather extensive coincidence at any $L$ found for
$N=2$ fermions between the Halperin/Laughlin wave function and one of the corresponding several RWM/FCI
0IE wave functions does not extend for larger, that is $N>2$, systems.

\section{Derivation of the analytic wave functions from the numerical FCI}
\label{fcian}

The exact numerical FCI wave functions are given by
\begin{align}
\Phi^{\rm CI} (z_1\sigma_1, \ldots , z_N\sigma_N) =
\sum_I c_{\rm CI}(I) {\cal D}_I(z_1\sigma_1, \ldots , z_N\sigma_N),
\label{phici}
\end{align}
with the basis Slater determinants that span the Hilbert space being 
\begin{align}
{\cal D}_I = {\rm Det}[ \psi_{l_r}(z_s)\sigma_{l_r}(s) ]/ \sqrt{N!}, 
\label{detexd}
\end{align}
where $r,s=1,\ldots,N$, the LLL single-particle space orbitals are given by Eq.\ (\ref{psilll})
(they are specified by the angular momentum index $l_r$),
and $\sigma$ signifies an up ($\alpha$) or a down ($\beta$) spin. The master index $I$ 
counts the number of ordered arrangements (lists) $\{j_1,j_2,\ldots,j_N\}$  under the 
restriction that $1 \leq j_1 < j_2 <\ldots < j_N \leq K$; $K \in \mathbb{N}$.

Next, one rewrites the CI wave functions $\Phi^{\rm CI}$ in 
Eq.~(\ref{phici}) as 
\begin{align}
\Phi^{\rm CI}_{\rm alg} (z_1\sigma_1, \ldots , z_N\sigma_N) =
\sum_I c_{\rm alg}(I) {\cal D}_I(z_1\sigma_1, \ldots , z_N\sigma_N),
\label{phialg1}
\end{align}
where the replacement of the subscript ``CI'' by ``alg'' corresponds to the fact that, using the 
symbolic language code, one obtains an equivalent  multivariate homogeneous integer polynomial 
$\Phi^{\rm CI}_{\rm alg}$ with algebraic coefficients $c_{\rm alg}$. In this way, our FCI
computer-assisted calculations, processed through the use of symbolic-language scripts, lead up to
analytical algebraic many-body wave-functions with direct correspondence to the numerical exact
diagonalization results. 

%------------------------------------------ begin table II -------
\begin{table}[t]
\caption{\label{tn3l2n1}
The numerical FCI coefficients, $c_{\rm CI}(I)$, in the CI expansion of the relative LLL {\it ground\/}
state, and the corresponding extracted algebraic ones, $c_{\rm alg}(I)$,
for $N=3$ fermions with total angular momentum $L=2$ and $S=S_z=1/2$. The spinful-fermion Slater
determinants ${\cal D}_I$ are specified through the set of single-particle angular momenta and spins,
$(l_1\uparrow,l_2\uparrow,l_3\downarrow)$.
}
\begin{ruledtabular}
\begin{tabular}{rcrc}
$I$ & $c_{\rm CI} (I)$ & $c_{\rm alg} (I)$ &
$(l_1\uparrow,l_2\uparrow,l_3\downarrow)$  \\ \hline
1 & -0.816496  & $ -\sqrt{2/3}  $ & (0,1,1) \\
2 &  0.577350  & $  1/\sqrt{3}  $ & (0,2,0) \\
\end{tabular}
\end{ruledtabular}
\end{table}
%------------------------------------------ end table II --------------------

%------------------------------------------ begin table III -------
\begin{table}[t]
\caption{\label{tn3l2n2}
The numerical FCI coefficients, $c_{\rm CI}(I)$, in the CI expansion of the
LLL {\it excited\/} state, and the corresponding extracted algebraic ones, $c_{\rm alg}(I)$,
for $N=3$ fermions with total angular momentum $L=2$ and $S=S_z=1/2$. The spinful-fermion Slater
determinants ${\cal D}_I$ are specified through the set of single-particle angular momenta and spins,
$(l_1\uparrow,l_2\uparrow,l_3\downarrow)$.
}
\begin{ruledtabular}
\begin{tabular}{rcrc}
$I$ & $c_{\rm CI} (I)$ & $c_{\rm alg} (I)$ &
$(l_1\uparrow,l_2\uparrow,l_3\downarrow)$  \\ \hline
1 & 0.577350 & $ 1/ \sqrt{3}  $ & (0,1,1) \\
2 & 0.816496 & $  \sqrt{2/3}  $ & (0,2,0) \\
\end{tabular}
\end{ruledtabular}
\end{table}
%------------------------------------------ end table III --------------------

Examples of such transcriptions of coefficients are given in Tables \ref{tn2l5n1}, \ref{tn3l2n1} and
\ref{tn3l2n2} corresponding to the two-fermion state $\Phi^{\rm CI}_{N=2}(L=5;\;1)$ at $L=5$ (see Fig.\
  \ref{spec32}), as well as the two three-fermion states $\Phi^{\rm CI}_{N=3}(L=2;\;1)$ and
$\Phi^{\rm CI}_{N=3}(L=2;\;2)$ at $L=2$ (see Fig.\ \ref{spec33}). We stress again that the polynomial
expressions produced through the expansion of $\Phi^{\rm CI}_{{N=2},{\rm alg}}(L=5;\;1)$,
$\Phi^{\rm CI}_{{N=3},{\rm alg}}(L=2;\;1)$, and $\Phi^{\rm CI}_{{N=3},{\rm alg}}(L=2;\;2)$ coincide with the
RWM polynomials in Eqs.\ (\ref{rwm2}), (\ref{rwm3l2n1}), and (\ref{rwm3l2n2}), respectively.

For the derivation of the $\Phi^{\rm CI}_{{N=3},{\rm alg}}$ analytic wave functions for $L=3$, see
Sec.\ \ref{a2}, and for those with $L=4$, see Appendix \ref{appb}. 

Validation of our closed-form analytic wave functions (see below) is achieved via direct comparison of
the full set of numerical CI coefficients, $c_{\rm CI}$, with those in $\Phi^{\rm CI}_{\rm alg}$ 
[Eq.~(\ref{phialg1})], thus circumventing uncertainties, associated with the common use of wave 
function overlap \cite{laug90,simochap,jainbook,palm20,yoso98}, due to the van Vleck-Anderson
orthogonality catastrophe \cite{vlec36,ande67,kohn99,deml12,ares18}.
Using symbolic scripts, we verify further that the fully-algebraic $\Phi^{\rm CI}_{\rm alg}$
[Eq.~(\ref{phialg1})] is indeed an eigenstate of the total spin, obeying the Fock condition 
\cite{fock40,luza11,jainbook,hald88}.

\section{Algebraic wave functions from FCI for $N=3$ and $L=3$}
\label{a2}

From Fig.\ \ref{spec33}, one sees that the LLL spectrum for $N=3$ fermions at $L=3$ consists of
4 states. The relative ground state has total spin $S=3/2$, whereas the three excited states have
total spin $S=1/2$. The FCI numerical solutions and the transcribed algebraic coefficients
$c_{\rm alg}$ are listed in Tables IV-VII.

%------------------------------------------ begin table IV -------
\begin{table}[t]
\caption{\label{tn3l3n1}
The numerical FCI coefficients, $c_{\rm CI}(I)$, in the CI expansion of the
LLL {\it ground\/} state, and the corresponding extracted algebraic ones, $c_{\rm alg}(I)$,
for $N=3$ fermions with total angular momentum $L=3$ and ($S=3/2$, $S_z=1/2$). The spinful-fermion Slater
determinants ${\cal D}_I$ here and in Tables \ref{tn3l3n2}-\ref{tn3l3n4} are specified through the set
of single-particle angular momenta and spins, $(l_1\uparrow,l_2\uparrow,l_3\downarrow)$.
}
\begin{ruledtabular}
\begin{tabular}{rcrc}
$I$ & $c_{\rm CI} (I)$ & $c_{\rm alg} (I)$ &
$(l_1\uparrow,l_2\uparrow,l_3\downarrow)$  \\ \hline
1 & -0.577350 & $ -1/\sqrt{3}  $ & (0,1,2) \\
2 & 0.577350  & $  1/\sqrt{3}  $ & (0,2,1) \\
3 & 0.0       & $  0           $ & (0,3,0) \\
4 & -0.577350 & $ -1/\sqrt{3}  $ & (1,2,0) \\
\end{tabular}
\end{ruledtabular}
\end{table}
%------------------------------------------ end table IV --------------------

Using Table \ref{tn3l3n1}, one finds for the algebraic transciption of the FCI relative ground state
\begin{align}
  &  \Phi^{\rm CI}_{N=3,{\rm alg}}(L=3;\;1) \propto \nonumber \\
  & (z_1-z_2)(z_1-z_3)(z_2-z_3)
  (\cz_{123}+\cz_{231}+\cz_{312})/\sqrt{3},
\label{n3l3n1} 
\end{align}
which coincides (within a proportionality constant) with the RWM wave function $\Phi^{\rm RWM}_{N=3}(L=3)$
given by Eq.\ (\ref{rwm3}) when $L=3$.

%------------------------------------------ begin table V -------
\begin{table}[t]
\caption{\label{tn3l3n2}
The numerical FCI coefficients, $c_{\rm CI}(I)$, in the CI expansion of the
LLL first excited state, and the corresponding extracted algebraic ones, $c_{\rm alg}(I)$,
for $N=3$ fermions with $L=3$ and ($S=1/2$, $S_z=1/2$).
}
\begin{ruledtabular}
\begin{tabular}{rcrc}
$I$ & $c_{\rm CI} (I)$ & $c_{\rm alg} (I)$ &
$(l_1\uparrow,l_2\uparrow,l_3\downarrow)$  \\ \hline
1 &  0.666666  &   2/3          & (0,1,2) \\
2 &  0.333333  &   1/3          & (0,2,1) \\
3 & -0.577350  & $-\sqrt{1/3}$  & (0,3,0) \\
4 & -0.333333  &  $-1/3$        & (1,2,0) \\
\end{tabular}
\end{ruledtabular}
\end{table}
%------------------------------------------ end table V --------------------

%------------------------------------------ begin table VI -------
\begin{table}[t]
\caption{\label{tn3l3n3}
The numerical FCI coefficients, $c_{\rm CI}(I)$, in the CI expansion of the
LLL second excited state, and the corresponding extracted algebraic ones, $c_{\rm alg}(I)$,
for $N=3$ fermions with $L=3$ and ($S=1/2$, $S_z=1/2$).
}
\begin{ruledtabular}
\begin{tabular}{rcrc}
$I$ & $c_{\rm CI} (I)$ & $c_{\rm alg} (I)$ &
$(l_1\uparrow,l_2\uparrow,l_3\downarrow)$  \\ \hline
1 & -0.333333  &  $-1/3$        & (0,1,2) \\      
2 &  0.333333  &   1/3          & (0,2,1) \\
3 & -0.577350  &  $-\sqrt{1/3}$ & (0,3,0) \\
4 &  0.666666  &   2/3          & (1,2,0) \\
\end{tabular}
\end{ruledtabular}
\end{table}
%------------------------------------------ end table VI --------------------

%------------------------------------------ begin table VII -------
\begin{table}[t]
\caption{\label{tn3l3n4}
The numerical FCI coefficients, $c_{\rm CI}(I)$, in the CI expansion of the
LLL third excited state, and the corresponding extracted algebraic ones, $c_{\rm alg}(I)$,
for $N=3$ fermions with $L=3$ and ($S=1/2$, $S_z=1/2$).
}
\begin{ruledtabular}
\begin{tabular}{rcrc}
$I$ & $c_{\rm CI} (I)$ & $c_{\rm alg} (I)$ &
$(l_1\uparrow,l_2\uparrow,l_3\downarrow)$  \\ \hline
1 &  0.333333   &  $1/3$        & (0,1,2) \\      
2 &  0.666666   &  $2/3$        & (0,2,1) \\
3 &  0.577350   &  $\sqrt{1/3}$ & (0,3,0) \\
4 &  0.333333   &  1/3          & (1,2,0) \\
\end{tabular}
\end{ruledtabular}
\end{table}
%------------------------------------------ end table VII --------------------

Using Table \ref{tn3l3n2}, one finds for the algebraic transciption of the FCI first excited state
\begin{align}
 &  \Phi^{\rm CI}_{N=3,{\rm alg}}(L=3;\;2) \propto \nonumber \\ 
 &  (z_1 + z_2 + z_3) \sum_{ijk} (z_i - z_j) (z_i + z_j - 2 z_k) \cz_{ijk},
\label{n3l3n2}
\end{align}
where the $ijk$ index sums over the cyclic permutations of \{1,2,3\}. One sees that
\begin{align}
  \Phi^{\rm CI}_{N=3,{\rm alg}}(L=3;\;2) \propto \Phi^{\rm RWM}_{N=3}(L=2) z_{\rm c.o.m.}^{N=3}.
  \label{rwmn3l3n2}
\end{align}
where the RWM wave function $\Phi^{\rm RWM}_{N=3}(L=2)$ is given by Eq.\ (\ref{rwm3}) when $L=2$.

%*********************** begin figure 5 *****************
\begin{figure*}[t]
\centering\includegraphics[width=13.0cm]{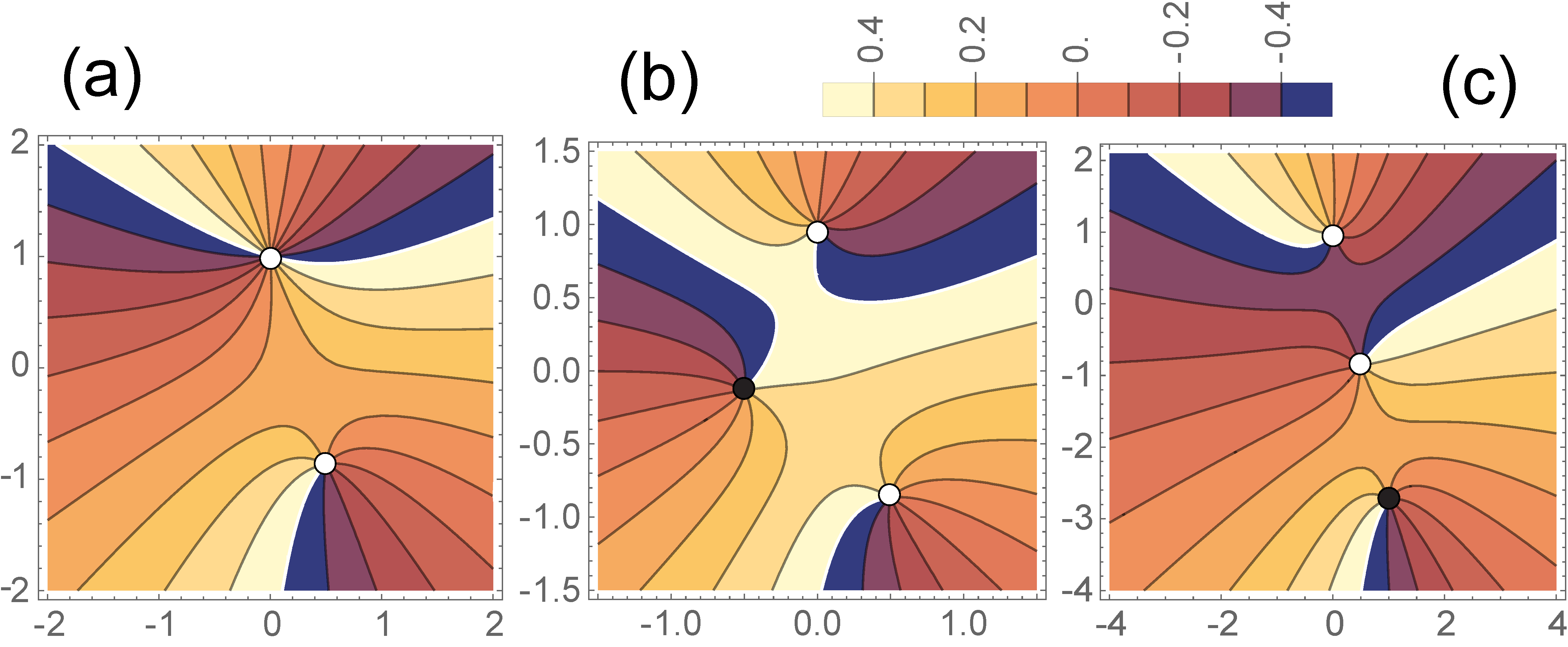}
\caption{%%%%
Maps of Arg[$\Upsilon^{N=3,L=4}_{(2,0,1)}(z_1)$], Arg[$\Phi^{\rm CI}_{N=3,{\rm alg}}(L=4;\;1)(z_1)$], and
Arg[$\Phi^{\rm CI}_{N=3,{\rm alg}}(L=4;\;2)(z_1)$] demonstrating the 
patterns of zeroes that are associated with the three different 0IE wave functions for $N=3$ and $L=4$.
(a) Case of the Halperin wave function given in Eq.\ (\ref{halp34}).
(b) Case of the FCI algebraic wave function with $(S=3/2,S_z=1/2)$ presented in Eq.\ (\ref{fcialg41}).
(c) Case of the FCI algebraic wave function with $(S=1/2,S_z=1/2)$ presented in Eq.\ (\ref{fcialg42}).
The two fixed fermions are at positions $z2=R e^{ \pi i/2}$ and $z3=R e^{-\pi i/3}$, with $R=1$..
The fixed zeroes are marked by a white dot. Zeroes not attached to $z_2$ or $z_3$ are marked by a black dot.
Lengths in units of $\Lambda$. The color scale is in units of $2\pi$; it is the same for all three panels.
}
\label{zeroes}
\end{figure*}
%*********************** end figure 5 *****************

Using Table \ref{tn3l3n3}, one finds for the algebraic transciption of the FCI second excited state
\begin{align}
  &  \Phi^{\rm CI}_{N=3,{\rm alg}}(L=3;\;3) \propto \nonumber \\
  & (z_1^2 + z_2^2 + z_3^2 - z_1 z_2 - z_1 z_3 - z_2 z_3) \sum_{ijk} (z_i - z_j) \cz_{ijk}.
\label{n3l3n3}
\end{align}
One sees that
\begin{align}
  \Phi^{\rm CI}_{N=3,{\rm alg}}(L=3;\;3) \propto \Phi^{\rm RWM}_{N=3}(L=1) Q_2,
  \label{rwmn3l3n3}
\end{align}
where $\Phi^{\rm RWM}_{N=3}(L=1)$ is given by Eq.\ (\ref{rwm3l1}),
and $Q_\lambda$ is a TI multipolar vibrational mode \cite{yann10,mott99,pape01} given by
\begin{align}
  Q_\lambda=\sum_{j=1}^N (z_j-z_{\rm c.o.m.})^\lambda.
\label{qlam}  
\end{align}

Using Table \ref{tn3l3n4}, one finds for the algebraic transciption of the FCI third excited state
\begin{align}
\Phi^{\rm CI}_{N=3,{\rm alg}}(L=3;\;4) \propto (z_1+z_2+z_3)^2 \sum_{ijk} (z_i - z_j) \cz_{ijk}.
\label{n3l3n4}
\end{align}
One sees that
\begin{align}
  \Phi^{\rm CI}_{N=3,{\rm alg}}(L=3;\;4) \propto \Phi^{\rm RWM}_{N=3}(L=1) (z^{N=3}_{\rm c.o.m})^2,  
  \label{rwmn3l3n4}
\end{align}
where the RWM wave function $\Phi^{\rm RWM}_{N=3}(L=1)$ is given by Eq.\ (\ref{rwm3}) when $L=1$.

\section{Contrast between the 0IE FCI wave functions and the trial Halperin wave function for $N=3$
 and $L=4$}
\label{halpp}

For spinful fermions, as a generalization of the Laughlin wave function \cite{laug90} (associated only
with the case of fully spin-polarized fermions), Halperin proposed \cite{halp83} the following expression
[denoted usually as $(p,q,r)$] \cite{halp83,girv96,tong16}
\begin{align}
\begin{split}
\Upsilon&_{(p,q,r)}(z,w)= \\
&\prod_{i<j}^{N_\uparrow} (z_i-z_j)^p \prod_{k<l}^{N_\downarrow} (w_k-w_l)^q
\prod_{i,k}^{N_\uparrow,N_\downarrow} (z_i-w_k)^r,
\end{split}
\label{halp}
\end{align}
where $p$, $q$, and $r$ are nonnegative intergers.

In Eq.\ (\ref{halp}), $z_i=r_i e^{i\theta_i}$ and $w_k=r_k e^{i\theta_k}$ are the space coordinates
(here in units of $\Lambda$) in the complex plane for the spin-up and spin-down fermions, respectively.
Note that the trivial Gaussian factors,
$\exp[-\sum_{i=1}^{N_\uparrow} z_i^* z_i/2]\exp[-\sum_{k=1}^{N_\downarrow} w_k^* w_k/2]$,
have been omitted in Eq.\ (\ref{halp}).

We note that the original Halperin proposal \cite{halp83} concerned only the case of equal spin-up and
spin-down electrons. Later, this proposal was generalized (see, e.g., Ref.\ \cite{tong16}) to the case
${N_\uparrow} \neq {N_\downarrow}$. We further note that, in general, expression (\ref{halp}) does not
honor the Fock condition \cite{fock40,luza11,jainbook,hald88},
and thus it does not conserve the total spin $S$,
a property that is a requirement for an assembly of a few trapped ultracold atoms.

It is apparent that expression (\ref{halp}) comprises only binary Jastrow-type $(z_i-z_j)^m$ factors, whereas
the FCI algebraic expressions comprise factots that are much more complex. Thus at best the agreement between
the $\Upsilon_{(p,q,r)}(z,w)$ Halperin wave functions and the exact $\Phi^{\rm CI}_{\rm alg}$ ones is limited
to a minority of 0IE cases. An example of such an agreement is given for two fermions by the 0IE states
$\Psi^{\rm RWM}_{N=2}(L)$ [see Eq.\ (\ref{rwm2})] which apparently agree with the $(0,0,L)$ Halperin
expressions, a fact that was championed in the experimental Ref.\ \cite{joch24}.
However, the case for $N=3$ fermions is not as supportive.

Indeed, an example of a noticeable disagreement appears for $N=3$ fermions with $S_z=1/2$ at $L=4$. In this
case,  according to Fig.\ \ref{spec33}, there are two singly degenerate 0IE FCI states, one with total spin
$S=3/2$ (indexed as No. 1) and the other with $S=1/2$ (indexed as No. 2). For these two FCI states, the
algebraic counterparts are given by (see Appendix \ref{appb} for the full derivation)
\begin{align}
 &  \Phi^{\rm CI}_{N=3,{\rm alg}}(L=4;\;1) \propto \nonumber \\
 &  (z_1 - z_2) (z_1 - z_3) (z_2 - z_3) (z_1 + z_2 + z_3) \sum_{ijk} \cz_{ijk},
\label{fcialg41}  
\end{align}  
and
\begin{align}
 &  \Phi^{\rm CI}_{N=3,{\rm alg}}(L=4;\;2) \propto \nonumber \\
 &  (z_1 - z_2) (z_1 - z_3) (z_2 - z_3) \sum_{ijk} (z_i + z_j-2 z_k) \cz_{ijk}.
\label{fcialg42}  
\end{align}  

On the other hand, the corresponding Halperin expression $(2,0,1)$ for $N=3$ with $S_z=1/2$ having $L=4$
[see Eq.\ (\ref{halp})] is given by
\begin{align}
  \Upsilon^{N=3,L=4}_{(2,0,1)}(z)=(z_1-z_2)^2 (z_1-z_3) (z_2-z_3),
  \label{halp34}
\end{align}
where we set $w_1=z_3$.

Expression (\ref{halp34}) strongly disagrees with both the FCI expressions
(\ref{fcialg41}) and (\ref{fcialg42}), a fact that can be further illustrated through the different behavior
of the wave function zeroes (see Fig.\ \ref{zeroes}); the zeroes are a component of the wave functions
employed in studies of highly correlated fermion systems (see Ref.\ \cite{laug90} and Sec.\ \ref{halpp}).
Indeed, the Halperin wave
function is characterized by a pattern consisting of a second-order zero (with each color encountered twice
when encircling the upper white dot in Fig.\ \ref{zeroes}, corresponding to a winding number of 2) 
and a single first-order one [see Fig.\ \ref{zeroes}(a)].
In contrast, the patterns of the FCI algebraic wave functions consist of three first-order zeroes
[see Figs.\ \ref{zeroes}(b) and \ref{zeroes}(c)]. 
Furthermore, expression (\ref{halp34}) violates the Fock conditions \cite{fock40,luza11} and thus it
does not conserve the total spin. We stress that total spin conservation
is an essential requirement for an assembly of ultarcold atoms. 

\section{Conclusions}
\label{concf}

Focusing here on systems of few spinful strongly-correlated ultracold fermionic $^6$Li atoms in a rapidly
rotating harmonic trap, which have been successfully realized in most recent experimental investigations
\cite{joch24}, we derive, using a newly-developed theoretical methodology that integrates 
exact numerical diagonalization using FCI with symbolic language processing, 
analytic (algebraic) total spin eigenstate wavefunctions that faithfully reproduce the full range of
numerically obtained results. We further show that the so-derived algebraic wave functions are
manifestations of collectively rotating and vibrating Wigner molecules, exhibiting unique spatial and
spectral quantum molecularization characteristics. The newly derived wavefunctions reach well beyond the
limited range of validity of previous Jastrow-type generalizations \cite{laug90,halp83}, thus providing
the impetus for future experiments targeting direct observations of the spatial, spin, and dynamical
correlated nature of hierarchically size-scalable larger ultracold atom systems in the spinful fractional
quantum Hall regime.

\textcolor{black}{
To reiterate our introductory comment, we trust that near-future experimental
investigations on rapidly-rotating ultra-cold 3-fermion systems, in addition to the case of two fermions
studied experimentally in Ref.\ \cite{joch24}, will conclusively validate the explicit correlated spinful
wave functions exposed here and the derivation methodology and Wigner molecule physics developed in this
work. Indeed, the most recent experimental detection \cite{joch24} of antipodal distributions of two
ultra-cold tweezer-held $^6$Li atoms are in agreement with the emergent physics uncovered here.
Furthermore, we anticipate that in line with the experimental developments, our current methodology could
be applied presently to larger sizes (of the order of 10 fermions), with much lager sizes requiring further
developments (perhaps employing machine-learning methods).
}

Finally, we remark that our theoretical methodology can be extended, as well, to bosonic systems under
current experimental investigations \cite{grei23}.

\acknowledgements

This work has been supported by a grant from the Air Force Office of Scientific Research (AFOSR)
under Grant No. FA9550-21-1-0198. Calculations were carried out at the GATECH Center for
Computational Materials Science.

\appendix

\section{BACKGROUND FOR THE $H_{\rm LLL}$ HAMILTONIAN [Eq.\ (\ref{hlll})]}
\label{appd}

{\color{black}
Requiring a very strong confinement of the harmonic trap 
along the axis of rotation ($\hbar \omega_{z} >> \hbar \omega_{\bot}$), 
freezes out the many body dynamics in the $z$-dimension, and the wavefunction
along this direction can be assumed to be permanently in the corresponding 
oscillator ground state.  
We are thus left with an effectively 2D system. For such a setup, 
the Hamiltonian for $N$ atoms of mass $M$ in a harmonic trap $(\omega_\bot)$ 
rotating at angular frequency $\Omega \hat{\bf z}$ is given by:  
\begin{equation}
H=\sum_{i=1}^N
\left( \frac{{\bf p}_{i}^{2}}{2M}+\frac{1}{2} 
M \omega_{\bot}^{2} {\bf r}_{i}^{2} \right) -\Omega {\cal L}
+\sum_{i < j}^{N} v({\bf r}_{i}-{\bf r}_{j}).
\label{3d_hamiltonian_1}
\end{equation}
Here ${\cal L}=-\hbar L=\sum_{i=1}^N \hat{{\bf z}}\!\cdot{\bf r}_i 
\times {\bf p}_{i}$ is the total angular momentum perpendicular to the $x-y$ plane;
$ {\bf r } \; =\; (x,y)$ and $ {\bf p } \; =( p_x,p_y )$ represent the single-particle position
and linear momentum in the $x-y$ plane, and $\omega_{\bot}$ is the frequency of the 2D harmonic trap.

The Hamiltonian can be rewritten in the form,
\begin{eqnarray}
H&=&\sum_{i=1}^N \left\{ \frac{\left({\bf p}_{i}-M\Omega\hat{\bf z} \times 
{\bf r}_{i}\right)^{2}}{2 M}+\frac{M}{2}(\omega_{\bot}^2-\Omega^2) {\bf r}_i^2
\right\} \nonumber \\ 
&&~~~~~~~~~~~+\sum_{i < j}^{N} v({\bf r}_{i}-{\bf r}_{j}). 
\label{3d_hamiltonian_3}
\end{eqnarray}
The kinetic part of this Hamiltonian is formally equivalent to that of the 
Hamiltonian in the symmetric gauge of an electron (of charge $e$ and mass
$m_e$) moving in two dimensions under a constant perpendicular magnetic field 
$B\hat{\bf z}$, if one makes the identification that the cyclotron frequency
$\omega_c=eB/(m_ec)\rightarrow 2\Omega$. 

We proceed to implement an LLL description of the few-body problem
by invoking two assumptions: (1) that the rotational frequency $\Omega$ is close to that of the
confining trap, i.e., $\Omega \rightarrow \omega_\bot$ (rapid rotation limit); in this case
the external confinement can be neglected in a first approximation and thus the single-particle
spectra correspond to the pure Landau problem and are organized in infinitely-degenerate Landau levels
that are separated by an energy gap of $2\hbar \omega_\bot$ (see Appendix A in Ref.\ \cite{yann07}), and (2)  
that the interaction strength is weak enough so that the mixing of 
Landau levels can be ignored. Since we work at zero temperature it then follows 
that all particles are in the lowest Landau level. 

Further, we can inquire about the influence of the trap frequency in this rapid rotation limit 
by employing the Fock-Darwin wave functions (see Appendix A in Ref.\ \cite{yann07})
to construct the single-particle basis for the FCI, and
by taking into account the fact that, as a result, the single-particle spectrum associated with the
Hamiltonian $H$ above [Eq.\ (\ref{3d_hamiltonian_1}) or Eq.\ (\ref{3d_hamiltonian_3})]
is given by $\epsilon^{\text{FD}}_{n,l}=\hbar [(2n+|l|+1) \omega_\bot - l \Omega]$ (Fock-Darwin spectrum
\cite{fock,darw,yann07}).  Then the 
restriction to the LLL requires $n=0$ (Fock-Darwin single-particle states with 
zero radial nodes) and this reduces the Hamiltonian $H$ above 
[Eq.\ (\ref{3d_hamiltonian_1}) or Eq.\ (\ref{3d_hamiltonian_3})] to the simpler form: 
\begin{equation}
H^\prime_{\text{LLL}}=N \hbar \omega_\bot + \hbar(\omega_{\bot}-\Omega) L
+\sum_{i < j}^{N} v({\bf r}_{i}-{\bf r}_{j}),
\label{hlll2}
\end{equation}
for which only the interaction term is non-trivial, since the many-body energy 
eigenstates are eigenstates of the total angular momentum as well; 
$L=\sum_{i=1}^N l_i$. Thus in Eq.\ (\ref{hlll}) of the main text we consider only
the interaction term.
}

\section{FCI ALGEBRAIC WAVE FUNCTIONS FOR THE FULL SPECTRUM FOR $N=2$}
\label{appa}

Here we present the algebraic wave functions for the remaining LLL states for $N=2$ and
$L=0-3$ whose energy spectrum was displayed in Fig.\ \ref{spec32}.
For $L=0-2$, we simply list the final result. For $L=3$, in addition, we present the FCI solutions and
the corresponding algebraic transcriptions, offering a further illustration of our methodology.

For $L=0$, this is a trivial case. The FCI solution consists of a single Slater determinant with
orbitals $(0\uparrow, 0\downarrow)$ and total spin ($S=0$, $S_z=0$). This yields for the space
part
\begin{align}
\Psi^{\rm CI}_{N=2,{\rm alg}}(L=0)=\Psi^{\rm RWM}_{N=2}(L=0)=1.
\end{align}  

For $L=1$, the spectrum consists of a triplet 0IE (No. 1) and a singlet excited state (No. 2) [see
Fig.\ \ref{spec32}]. One has:
\begin{align}
\Psi^{\rm CI}_{N=2,{\rm alg}}(L=1;\;1) \propto (z_1-z_2) = \Psi^{\rm RWM}_{N=2}(L=1)
\end{align}
and
\begin{align}
\Psi^{\rm CI}_{N=2,{\rm alg}}(L=1;\;2) \propto (z_1+z_2) \propto z_{\rm c.o.m.}^{N=2}.
\end{align}

For $L=2$, the spectrum consists of a singlet 0IE (No. 1), a triplet 0IE state (No. 2), and
a singlet excited state (No. 3) [see Fig.\ \ref{spec32}]. One has:
\begin{align}
\Psi^{\rm CI}_{N=2,{\rm alg}}(L=2;\;1) \propto (z_1-z_2)^2 = \Psi^{\rm RWM}_{N=2}(L=2),
\end{align}

\begin{align}
&  \Psi^{\rm CI}_{N=2,{\rm alg}}(L=2;\;2) \propto (z_1-z_2)(z_1+z_2) \propto \nonumber \\
&  \Psi^{\rm RWM}_{N=2}(L=1) z_{\rm c.o.m.}^{N=2},
\end{align}
and
\begin{align}
  \Psi^{\rm CI}_{N=2,{\rm alg}}(L=2;\;3) \propto (z_1+z_2)^2 \propto
  (z_{\rm c.o.m.}^{N=2})^2,
\end{align}

%------------------------------------------ begin table VIII -------
\begin{table}[t]
\caption{\label{tn2l3n1}
The numerical FCI coefficients, $c_{\rm CI}(I)$, in the CI expansion of the first 0IE LLL state (a
triplet), and the corresponding extracted algebraic ones, $c_{\rm alg}(I)$, for $N=2$ fermions with
total angular momentum $L=3$ and spin ($S=1$, $S_z=0$); see Fig.\ \ref{spec32}.
The Slater determinants ${\cal D}_I$ are specified through the set of single-particle angular momenta
and spins $(l_1\uparrow,l_2\downarrow)$.
}
\begin{ruledtabular}
\begin{tabular}{rccc}
$I$ & $c_{\rm CI} (I)$ & $c_{\rm alg} (I)$ & $(l_1\uparrow,l_2\downarrow)$  \\ \hline
1 &  0.353553  &  $  \sqrt{1/8} $  & (0,3) \\
2 & -0.612372  &  $ -\sqrt{3/8} $  & (1,2) \\
3 &  0.612372  &  $  \sqrt{3/8} $  & (2,1) \\
4 & -0.353553  &  $ -\sqrt{1/8} $  & (3,0) \\
\end{tabular}
\end{ruledtabular}
\end{table}
%------------------------------------------ end table VIII --------------------

Using Table \ref{tn2l3n1}, one finds for the algebraic transcription of the space part of the
first FCI 0IE state with total angular momentum $L=3$ and spin ($S=1$, $S_z=0$) [see Fig.\ \ref{spec32}]:
\begin{align}
  \Psi^{\rm CI}_{N=2,{\rm alg}}(L=3;\;1) \propto (z_1-z_2)^3 = \Psi^{\rm RWM}_{N=2}(L=3),
\label{cin2l3n1} 
\end{align}
where $\Psi^{\rm RWM}_{N=2}(L)$ is given by Eq.\ (\ref{rwm2}).

%------------------------------------------ begin table IX -------
\begin{table}[t]
\caption{\label{tn2l3n2}
Same as in Table \ref{tn2l3n1}, but for the second 0IE state (a triplet) with $N=2$, $L=3$,
and ($S=1$, $S_z=0$); see Fig.\ \ref{spec32}.   
}
\begin{ruledtabular}
\begin{tabular}{rccc}
$I$ & $c_{\rm CI} (I)$ & $c_{\rm alg} (I)$ & $(l_1\uparrow,l_2\downarrow)$  \\ \hline
1 & -0.612372  &  $ -\sqrt{3/8} $  & (0,3) \\
2 & -0.353553  &  $ -\sqrt{1/8} $  & (1,2) \\
3 &  0.353553  &  $  \sqrt{1/8} $  & (2,1) \\
4 &  0.612372  &  $  \sqrt{3/8} $  & (3,0) \\
\end{tabular}
\end{ruledtabular}
\end{table}
%------------------------------------------ end table IX --------------------

Using Table \ref{tn2l3n2}, one finds for the algebraic transcription of the space part of the
second FCI 0IE state with total angular momentum $L=3$ and spin ($S=1$, $S_z=0$) [see Fig.\ \ref{spec32}]: 
\begin{align}
  & \Psi^{\rm CI}_{N=2,{\rm alg}}(L=3;\;2) \propto (z_1-z_2) (z_1+z_2)^2 \propto \nonumber \\ 
  & \Psi^{\rm RWM}_{N=2}(L=1) (z_{\rm c.o.m.}^{N=2})^2,
\label{cin2l3n2} 
\end{align}
where $\Psi^{\rm RWM}_{N=2}(L)$ is given by Eq.\ (\ref{rwm2}).

%------------------------------------------ begin table X -------
\begin{table}[t]
\caption{\label{tn2l3n3}
Same as in Table \ref{tn2l3n1}, but for the third 0IE state (a singlet) with $N=2$, $L=3$,
and ($S=0$, $S_z=0$); see Fig.\ \ref{spec32}.   
}
\begin{ruledtabular}
\begin{tabular}{rccc}
$I$ & $c_{\rm CI} (I)$ & $c_{\rm alg} (I)$ & $(l_1\uparrow,l_2\downarrow)$  \\ \hline
1 &  0.612372 &  $ \sqrt{3/8}  $  & (0,3) \\
2 & -0.353553 &  $ -\sqrt{1/8} $  & (1,2) \\
3 & -0.353553 &  $ -\sqrt{1/8} $  & (2,1) \\
4 &  0.612372 &  $  \sqrt{3/8} $  & (3,0) \\
\end{tabular}
\end{ruledtabular}
\end{table}
%------------------------------------------ end table X --------------------

Using Table \ref{tn2l3n3}, one finds for the algebraic transcription of the space part of the
third FCI 0IE state with total angular momentum $L=3$ and spin ($S=0$, $S_z=0$) [see Fig.\ \ref{spec32}]: 
\begin{align}
  & \Psi^{\rm CI}_{N=2,{\rm alg}}(L=3;\;3) \propto (z_1-z_2)^2 (z_1+z_2) \propto \nonumber \\ 
  & \Psi^{\rm RWM}_{N=2}(L=2) z_{\rm c.o.m.}^{N=2},
\label{cin2l3n3} 
\end{align}
where $\Psi^{\rm RWM}_{N=2}(L)$ is given by Eq.\ (\ref{rwm2}).

%------------------------------------------ begin table XI -------
\begin{table}[t]
\caption{\label{tn2l3n4}
Same as in Table \ref{tn2l3n1}, but for the excited state (a singlet with energy 0.5 $g/(\pi\Lambda^2)$
with $N=2$, $L=3$, and ($S=0$, $S_z=0$); see Fig.\ \ref{spec32}.
}
\begin{ruledtabular}
\begin{tabular}{rccc}
$I$ & $c_{\rm CI} (I)$ & $c_{\rm alg} (I)$ & $(l_1\uparrow,l_2\downarrow)$  \\ \hline
1 &  0.353553 &  $ \sqrt{1/8} $  & (0,3) \\
2 &  0.612372 &  $ \sqrt{3/8} $  & (1,2) \\
3 &  0.612372 &  $ \sqrt{3/8} $  & (2,1) \\
4 &  0.353553 &  $ \sqrt{1/8} $  & (3,0) \\
\end{tabular}
\end{ruledtabular}
\end{table}
%------------------------------------------ end table XI --------------------

Using Table \ref{tn2l3n4}, one finds for the algebraic transcription of the space part of the
excited FCI state with energy 0.5 $g/(\pi\Lambda^2)$, total angular momentum $L=3$ and spin
($S=0$, $S_z=0$) [see Fig.\ \ref{spec32}]: 
\begin{align}
\Psi^{\rm CI}_{N=2,{\rm alg}}(L=3;\;4) \propto (z_1+z_2)^3 \propto (z_{\rm c.o.m.}^{N=2})^3.
\label{cin2l3n4} 
\end{align}

\section{ALGEBRAIC WAVE FUNCTIONS FROM FCI FOR $N=3$ AND $L=4$}
\label{appb}

%------------------------------------------ begin table XII -------
\begin{table}[t]
\caption{\label{tn3l4n1}
The numerical FCI coefficients, $c_{\rm CI}(I)$, and the corresponding extracted algebraic ones,
$c_{\rm alg}(I)$, in the CI expansion of the 0IE LLL state for $N=3$ fermions with total angular
momentum $L=4$ and $S=S_z=1/2$ [No. 1, see Fig.\ 2]. The spinful-fermion Slater
determinants ${\cal D}_I$ are specified through the set of single-particle angular momenta and spins,
$(l_1\uparrow,l_2\uparrow,l_3\downarrow)$.
}
\begin{ruledtabular}
\begin{tabular}{rcrc}
$I$ & $c_{\rm CI} (I)$ & $c_{\rm alg} (I)$ &
$(l_1\uparrow,l_2\uparrow,l_3\downarrow)$  \\ \hline
1 &  0.516401   & $ \sqrt{4/15}   $ & (0,1,3) \\
2 & -0.632435   & $ -\sqrt{2/5}   $ & (0,2,2) \\
3 &  0.258131   & $ \sqrt{1/15}   $ & (0,3,1) \\
4 &  0.0000     & $ 0             $ & (0,4,0) \\
5 &  0.447269   & $ \sqrt{1/5}    $ & (1,2,1) \\
6 & -0.258211   & $ -\sqrt{1/15}  $ & (1,3,0) \\
\end{tabular}
\end{ruledtabular}
\end{table}
%------------------------------------------ end table XII --------------------
% 1 3  4  0.51640120 -0.63243594  0.25813142  0.00004002  0.44726905 -0.25821144

Using Table \ref{tn3l4n1}, one finds for the algebraic transcription of the 0IE 
FCI state with total angular momentum $L=4$ and spin $S=S_z=1/2$ [see Fig.\ 2]:
\begin{align}
 & \Phi^{\rm CI}_{N=3,{\rm alg}}(L=4;\;1) \propto (z_1-z_2)(z_1-z_3)(z_2-z_3)\times \nonumber \\
 & \sum_{ijk}(z_i+z_j-2 z_k) \cz_{ijk}\propto \nonumber \\
 & \sum_{ijk} \Psi^{\rm RWM}_{ijk}(L=3) (z_k-z_{\rm c.o.m.}^{N=3}) \cz_{ijk}.
\label{cin3l4n1}
\end{align}

%------------------------------------------ begin table XIII -------
\begin{table}[t]
\caption{\label{tn3l4n2}
The numerical FCI coefficients, $c_{\rm CI}(I)$, and the corresponding extracted algebraic ones,
$c_{\rm alg}(I)$, in the CI expansion of the 0IE LLL state for $N=3$ fermions with total angular
momentum $L=4$ and ($S=3/2$, $S_z=1/2$) [No. 2, see Fig.\ 2]. The spinful-fermion Slater
determinants ${\cal D}_I$ are specified through the set of single-particle angular momenta and spins,
$(l_1\uparrow,l_2\uparrow,l_3\downarrow)$.
}
\begin{ruledtabular}
\begin{tabular}{rcrc}
$I$ & $c_{\rm CI} (I)$ & $c_{\rm alg} (I)$ &
$(l_1\uparrow,l_2\uparrow,l_3\downarrow)$  \\ \hline
1 &  0.577332  & $  \sqrt{1/3}   $ & (0,1,3) \\
2 &  0.0000    & $ 0             $ & (0,2,2) \\
3 & -0.577358  & $ -\sqrt{1/3}   $ & (0,3,1) \\
4 &  0.0000    & $ 0             $ & (0,4,0) \\
5 &  0.0000    & $ 0             $ & (1,2,1) \\
6 &  0.577358  & $  \sqrt{1/3}   $ & (1,3,0) \\
\end{tabular}
\end{ruledtabular}
\end{table}
%------------------------------------------ end table XIII --------------------
% 2 3  4  0.57733287  0.00002131 -0.57735897 -0.00000002 -0.00001505  0.57735897

Using Table \ref{tn3l4n2}, one finds for the algebraic transcription of the 0IE FCI state with total
angular momentum $L=4$ and spin ($S=3/2$, $S_z=1/2$) [see Fig.\ 2]:
\begin{align}
 & \Phi^{\rm CI}_{N=3,{\rm alg}}(L=4;\;2) \propto (z_1-z_2)(z_1-z_3)(z_2-z_3) \times \nonumber \\
 & (z_1+z_2+z_3)  \sum_{ijk} \cz_{ijk}  \propto \nonumber \\
 & \Phi^{\rm RWM}_{N=3}(L=3) z_{\rm c.o.m.}^{N=3}.
\label{cin3l4n2}
\end{align}

%------------------------------------------ begin table XIV -------
\begin{table}[t]
\caption{\label{tn3l4n3}
The numerical FCI coefficients, $c_{\rm CI}(I)$, and the corresponding extracted algebraic ones,
$c_{\rm alg}(I)$, in the CI expansion of the first non-zero LLL state for $N=3$ fermions
with total angular momentum $L=4$ and $S=S_z=1/2$ [No. 3, see Fig.\ 2]. The
spinful-fermion Slater determinants ${\cal D}_I$ are specified through the set of single-particle
angular momenta and spins, $(l_1\uparrow,l_2\uparrow,l_3\downarrow)$.
}
\begin{ruledtabular}
\begin{tabular}{rcrc}
$I$ & $c_{\rm CI} (I)$ & $c_{\rm alg} (I)$ &
$(l_1\uparrow,l_2\uparrow,l_3\downarrow)$  \\ \hline
1 &  0.471404  & $  \sqrt{2/9}   $ & (0,1,3) \\
2 &  0.577350  & $  \sqrt{1/3}   $ & (0,2,2) \\
3 &  0.0000    & $  0            $ & (0,3,1) \\
4 & -0.471404  & $ -\sqrt{2/9}   $ & (0,4,0) \\
5 &  0.0000    & $  0            $ & (1,2,1) \\
6 & -0.471404  & $ -\sqrt{2/9}   $ & (1,3,0) \\
\end{tabular}
\end{ruledtabular}
\end{table}
%------------------------------------------ end table XIV --------------------
% 3 3  4  0.47140449  0.57735025  0.00000004 -0.47140462  0.00000005 -0.47140448

Using Table \ref{tn3l4n3}, one finds for the algebraic transcription of the first non-zero FCI state
with total angular momentum $L=4$ and spin $S=S_z=1/2$ [see Fig.\ 2]:
\begin{align}
  & \Phi^{\rm CI}_{N=3,{\rm alg}}(L=4;\;3) \propto (z_1+z_2+z_3)^2 \times \nonumber \\
  & \sum_{ijk} (z_i-z_j)(z_i+z_j-2z_k) \cz_{ijk}  \propto \nonumber \\
  & \Phi^{\rm RWM}_{N=3}(L=2) (z_{\rm c.o.m.}^{N=3})^2.
\label{cin3l4n3}
\end{align}

%------------------------------------------ begin table XV -------
\begin{table}[t]
\caption{\label{tn3l4n4}
The numerical FCI coefficients, $c_{\rm CI}(I)$, and the corresponding extracted algebraic ones,
$c_{\rm alg}(I)$, in the CI expansion of the second non-zero (excited) LLL state for $N=3$ fermions
with total angular momentum $L=4$ and $S=S_z=1/2$ [No. 4, see Fig.\ 2]. The
spinful-fermion Slater determinants ${\cal D}_I$ are specified through the set of single-particle
angular momenta and spins, $(l_1\uparrow,l_2\uparrow,l_3\downarrow)$.
}
\begin{ruledtabular}
\begin{tabular}{rcrc}
$I$ & $c_{\rm CI} (I)$ & $c_{\rm alg} (I)$ &
$(l_1\uparrow,l_2\uparrow,l_3\downarrow)$  \\ \hline
1 & -0.172132   & $ -\sqrt{4/135}  $ & (0,1,3) \\
2 &  0.210818   & $  \sqrt{2/45}   $ & (0,2,2) \\
3 & -0.516397   & $ -\sqrt{36/135} $ & (0,3,1) \\
4 &  0.430331   & $  \sqrt{5/27}   $ & (0,4,0) \\
5 &  0.596284   & $  \sqrt{48/135} $ & (1,2,1) \\
6 & -0.344265   & $ -\sqrt{16/135} $ & (1,3,0) \\
\end{tabular}
\end{ruledtabular}
\end{table}
%------------------------------------------ end table XV --------------------
%  4 3  4 -0.17213255  0.21081854 -0.51639780  0.43033153  0.59628471 -0.34426525

Using Table \ref{tn3l4n4}, one finds for the algebraic transcription of the second non-zero FCI state
with total angular momentum $L=4$ and spin $S=S_z=1/2$ [see Fig.\ 2]:
\begin{align}
  & \Phi^{\rm CI}_{N=3,{\rm alg}}(L=4;\;4) \propto \sum_{ijk}(z_i-z_j)(z_i+z_j-2z_k) \times \nonumber \\
  & (5z_i^2 - 8 z_i z_j + 5 z_j^2 - 2 z_i z_k - 2 z_j z_k + 2 z_k^2) \cz_{ijk}  \propto \nonumber \\
  & 12 \Phi^{\rm RWM}_{N=3}(L=2) Q_2 -3 \Phi^{\rm RWM}_{N=3}(L=4).
\label{cin3l4n4}
\end{align}

We note that the wave function $\Phi^{\rm CI}_{N=3,{\rm alg}}(L=4;\;4)$ is translationally
invariant; see the red arrow in Fig.\ 2.

%------------------------------------------ begin table XVI -------
\begin{table}[t]
\caption{\label{tn3l4n5}
The numerical FCI coefficients, $c_{\rm CI}(I)$, and the corresponding extracted algebraic ones,
$c_{\rm alg}(I)$, in the CI expansion of the third non-zero (excited) LLL state for $N=3$ fermions
with total angular momentum $L=4$ and $S=S_z=1/2$ [No. 5, see Fig.\ 2]. The
spinful-fermion Slater determinants ${\cal D}_I$ are specified through the set of single-particle
angular momenta and spins, $(l_1\uparrow,l_2\uparrow,l_3\downarrow)$.
}
\begin{ruledtabular}
\begin{tabular}{rcrc}
$I$ & $c_{\rm CI} (I)$ & $c_{\rm alg} (I)$ &
$(l_1\uparrow,l_2\uparrow,l_3\downarrow)$  \\ \hline
1 & -0.333333  & $-1/3$        & (0,1,3) \\
2 &  0.0000    & 0             & (0,2,2) \\
3 &  0.0000    & 0             & (0,3,1) \\
4 & -0.666666  & $-2/3$        & (0,4,0) \\
5 &  0.577350  & $\sqrt{1/3}$  & (1,2,1) \\
6 &  0.333333  & $1/3 $        & (1,3,0) \\
\end{tabular}
\end{ruledtabular}
\end{table}
%------------------------------------------ end table XVI --------------------
% 5 3  4 -0.33333342 -0.00000002 -0.00000003 -0.66666655  0.57735034  0.33333335

Using Table \ref{tn3l4n5}, one finds for the algebraic transcription of the third non-zero FCI state
with total angular momentum $L=4$ and spin $S=S_z=1/2$ [see Fig.\ 2]:
\begin{align}
  & \Phi^{\rm CI}_{N=3,{\rm alg}}(L=4;\;5) \propto \nonumber \\
  & (z_1 + z_2 + z_3) (z_1^2 - z_1 z_2 + z_2^2 - z_1 z_3 - z_2 z_3 + z_3^2) \times \nonumber \\
  & \sum_{ijk} (z_i-z_j) \cz_{ijk}  \propto \nonumber \\
  & \Phi^{\rm RWM}_{N=3}(L=1) Q_2 z^{N=3}_{\rm c.o.m.}. 
\label{cin3l4n5}
\end{align}

%------------------------------------------ begin table XVII -------
\begin{table}[t]
\caption{\label{tn3l4n6}
The numerical FCI coefficients, $c_{\rm CI}(I)$, and the corresponding extracted algebraic ones,
$c_{\rm alg}(I)$, in the CI expansion of the fourth non-zero (excited) LLL state for $N=3$ fermions
with total angular momentum $L=4$ and $S=S_z=1/2$ [No. 6, see Fig.\ 2]. The
spinful-fermion Slater determinants ${\cal D}_I$ are specified through the set of single-particle
angular momenta and spins, $(l_1\uparrow,l_2\uparrow,l_3\downarrow)$.
}
\begin{ruledtabular}
\begin{tabular}{rcrc}
$I$ & $c_{\rm CI} (I)$ & $c_{\rm alg} (I)$ &
$(l_1\uparrow,l_2\uparrow,l_3\downarrow)$  \\ \hline
1 &  0.192450  & $ \sqrt{1/27} $    & (0,1,3) \\
2 &  0.471404  & $ \sqrt{2/9}  $    & (0,2,2) \\
3 &  0.577350  & $ \sqrt{1/3}  $    & (0,3,1) \\
4 &  0.384900  & $ \sqrt{4/27} $    & (0,4,0) \\
5 &  0.333333  & $ 1/3         $    & (1,2,1) \\
6 &  0.384900  & $ \sqrt{4/27} $    & (1,3,0) \\
\end{tabular}
\end{ruledtabular}
\end{table}
%------------------------------------------ end table XVII --------------------
% 6 3  4  0.19245009  0.47140452  0.57735026  0.38490021  0.33333333  0.38490016

Using Table \ref{tn3l4n6}, one finds for the algebraic transcription of the fourth non-zero FCI state
with total angular momentum $L=4$ and spin $S=S_z=1/2$ [see Fig.\ 2]:
\begin{align}
  & \Phi^{\rm CI}_{N=3,{\rm alg}}(L=4;\;6) \propto \nonumber \\
  & (z_1 + z_2 + z_3)^3 \sum_{ijk} (z_i-z_j) \cz_{ijk} \propto \nonumber \\
  & \Phi^{\rm RWM}_{N=3}(L=1) (z^{N=3}_{\rm c.o.m.})^3. 
\label{cin3l4n6}
\end{align}

\section{BRIEF NOTE ON THE CONFIGURATION INTERACTION METHOD AS EMPLOYED HERE}
\label{appc}

The full configuration interaction (FCI) methodology has a long history, starting in quantum 
chemistry; see Refs.\ \cite{shav98,szabo}. The method was adapted to two dimensional problems and 
found extensive applications in the fields of semiconductor quantum dots, ultracold atoms in harmonic traps,
and moir\'e materials
\cite{yann03,szaf03,yann04,ront06,yann06.3,yann07,yann07.2,yann07.3,yann09,yann22.2,yann22.3},
as well as in the field of the fractional quantum Hall effect for fermions
\cite{yann03,yann04,yann20,yann21,yann06.3} and bosons \cite{yann07.2}.

Our 2D adaptation of the FCI in the case of ultracold atoms in the LLL was described earlier in
Refs.\ \cite{yann20,yann21}.
For fermions, the pivotal step of the FCI is the determination of the coefficients in the
multi-determinantal expansion described by Eq.\ (\ref{phici}). The LLL single-particle space
orbitals that are employed in the building of the single-particle basis used to construct the Slater
determinants ${\cal D}_I$, which span the many-body Hilbert space, are given by Eq.\ (\ref{psilll}).

Next, one diagonalizes the associated Hamiltonian matrix by calculating the matrix elements
$\langle {\cal D}_I |H_{\rm LLL}| {\cal D}_J \rangle$ [for the $H_{\rm LLL}$, see Eq.\ (\ref{hlll})]
through the application of the Slater-Condon rules \cite{slat29,cond30} and using the fact that the two-body
matrix elements of the 2D delta contact interaction are given by
\begin{align}
\frac{1}{\pi}\frac{\delta_{l_1+l_2,l_3+l_4}}{\sqrt{l_1! l_2! l_3! l_4!}}
\frac{(l_1+l_2)!}{2^{l_1+l_2+1}}.
\label{dltme}
\end{align}

We note that the equivalence between an applied perpendicular magnetic field, $B$, and the
rotational frequency, $\omega_\bot$, as well as the structure and expressions for the LLL orbitals in both cases are discussed in detail in Appendix A of Ref.\ \cite{yann07}.

We further note that for the CI diagonalization, a small perturbing term $V_P$ (e.g., a small hard-wall
boundary \cite{maca17,yann21} or a small Coulombic term) needs to be added to the LLL
Hamiltonian $H_{\rm LLL}$. This has a negligible influence on the numerical eigenvalues,
but it is instrumental in lifting the degeneracies among the zero-energy states, and thus in producing
numerical FCI states whose total spin $S$ is a good quantum number.

\nocite{*}
\bibliographystyle{apsrev4-2}
\bibliography{mycontrols,three_two_rotating_fermions_LLL_pra_r}

\end{document}